\documentclass[trackchanges,twocolumn]{aastex701}
\usepackage{CJKutf8}

\begin{document}
\begin{CJK*}{UTF8}{gbsn}

\title{The Merian Survey: A Statistical Census of Bright Satellites of Sub-Milky Way-Mass Galaxies}

\author[orcid=0000-0002-4858-0396,sname='Mandacar\'u Guerra']{Lucas S. Mandacar\'u Guerra}
\affiliation{Department of Physics, Princeton University, Princeton, NJ 08544, USA}
\affiliation{Joseph Henry Laboratories and Leinweber Forum for Theoretical Physics, Princeton University, Princeton, NJ 08544, USA}
\email{lm4604@princeton.edu}

\author[0000-0001-9592-4190]{Jiaxuan Li (李嘉轩)}
\affil{Department of Astrophysical Sciences, Princeton University, 4 Ivy Lane, Princeton, NJ 08544, USA}
\affiliation{Kavli Institute for Particle Astrophysics and Cosmology, Stanford University, Stanford, CA 94305, USA}
\email{jiaxuanl@stanford.edu}

\author[0000-0002-7922-9726]{Yue Pan (潘越)}
\affil{Department of Astrophysical Sciences, Princeton University, 4 Ivy Lane, Princeton, NJ 08544, USA}
\email{yue.pan@princeton.edu}

\author[0000-0002-5612-3427]{Jenny E. Greene}
\affil{Department of Astrophysical Sciences, Princeton University, 4 Ivy Lane, Princeton, NJ 08544, USA}
\email{jgreene@astro.princeton.edu}

\author[0000-0002-1841-2252]{Shany Danieli}
\affil{School of Physics and Astronomy, Tel Aviv University, Tel Aviv 69978, Israel}
\email{sdanieli@tauex.tau.ac.il}

\author[0000-0001-6442-5786]{Joy Bhattacharyya}
\affiliation{Department of Physics \& Astronomy, Amherst College, 6 East Drive, Amherst, MA 01002, USA}
\email{sdanieli@tauex.tau.ac.il}

\author[0000-0002-0372-3736]{Alyson M. Brooks}
\affiliation{Department of Physics \& Astronomy, Rutgers, the State University of New Jersey, Piscataway, NJ 08854, USA}
\email{jbhattacharyya@amherst.edu}

\author[0000-0003-1385-7591]{Song Huang (黄崧)}
\affil{Department of Astronomy, Tsinghua University, Beijing 100084, China}
\email{shuang@tsinghua.edu.cn}

\author[0000-0002-3677-3617]{Alexie Leauthaud}
\affil{Department of Astronomy and Astrophysics, University of California, Santa Cruz, 1156 High Street, Santa Cruz, CA 95064, USA}
\email{alexie@ucsc.edu}

\author[0000-0001-7729-6629]{Yifei Luo}
\affil{Lawrence Berkeley National Laboratory, 1 Cyclotron Road, Berkeley, CA 94720, USA}
\email{yifeiluo@lbl.gov}

\author[0000-0002-8040-6785]{Annika H. G. Peter}
\affiliation{Department of Physics, The Ohio State University, 191 W. Woodruff Ave., Columbus OH 43210, USA}
\affiliation{Center for Cosmology and Astroparticle Physics, The Ohio State University, 191 W. Woodruff Ave., Columbus OH 43210, USA}
\affiliation{Department of Astronomy, The Ohio State University, 140 W. 18th Ave., Columbus OH 43210, USA}
\email{peter.33@osu.edu}

\author[0000-0002-2897-6326]{Conghao Zhou}
\affil{Department of Astronomy and Astrophysics, University of California, Santa Cruz, 1156 High Street, Santa Cruz, CA 95064, USA}
\email{zhou.conghao@ucsc.edu}

\begin{abstract}
Satellite populations provide sensitive tests of the galaxy-halo connection, but studies have largely focused on Milky Way~(MW)-mass host galaxies or a small number of nearby dwarf galaxy hosts. Here we present a statistical census of bright, star-forming satellite galaxies around sub-MW-mass hosts using the first data release of the Merian Survey. Our sample consists of 357 hosts with stellar mass $10^{9.5}<M_{\star,\text{host}}<10^{10.5} \ \text{M}_\odot$ at redshifts $0.07<z<0.09$, all selected to be central galaxies. Merian's medium-band imaging isolates H$\alpha$ and [O~\textsc{iii}] emission lines for galaxies in this redshift range, enabling efficient selection of star-forming satellites down to $M_{\star,\text{sat}}\gtrsim 10^{7.5} \,\text{M}_\odot$. We identify 471 satellite candidates within the projected virial radii of these hosts. After statistically subtracting the contribution from foreground and background galaxies, we find $148\pm 7$ satellite candidates associated with the host sample. We find that $\mathord{\sim} 60\%$ of hosts have $N_{\rm sat}< 1$ massive satellites within their virial radii, $\mathord{\sim} 22\%$ have $1\leq N_{\rm sat}< 2$, $\mathord{\sim} 10\%$ have $2\leq N_{\rm sat}< 3$, and $\mathord{\sim} 8\%$ have $N_{\rm sat} > 3$. Satellite abundance increases with host stellar mass across the sub-MW regime, extending the observed trend from dwarf to MW-mass galaxies and broadly agreeing with predictions from the TNG50 simulation. The radial distribution of satellites around sub-MW-mass hosts is similar to that around MW-mass hosts and broadly consistent with TNG50. Together, these results extend our empirical view of satellite populations across the previously sparsely sampled transition between dwarf and MW-mass hosts.

\end{abstract}

\section{Introduction} 

Satellite dwarf galaxies represent rigorous tests of models of galaxy formation and evolution. The number, distribution, and properties of satellite galaxies have posed several challenges to Lambda Cold Dark Matter ($\Lambda$CDM) cosmology, refining our understanding of structure formation~\citep{Moore.etal.1999, Bullock2017, Sales2022}. Satellites have also motivated the study of alternative dark matter models~\citep{Spergel2000,Nadler2021,Kim2021, Gemmel2024,Esteban2024}, inflation~\citep{Esteban2024,Dekker2025}, stellar feedback~\citep{Hopkins2011}, and the evolution of host galaxies through processes such as tidal stripping and mergers~\citep{Zolotov2012,Brooks2013,Brooks2014,Fillingham.etal.2018, Wechsler.Tinker.2018, Joshi.etal.2025}. To obtain a detailed understanding of satellite populations, we require large statistical samples across different environments. 

Past studies have focused on the satellites of the Milky Way (MW) and M31~\citep[e.g.,][]{McConnachie.2012, Savino2025,Geha2026}. The detailed characterization of these populations has revealed great diversity in properties such as star formation history~\citep{Weisz2013} and gas content~\citep{Putman2012}. Early studies extended such searches to individual nearby MW analogs spanning a range of environments \citep{Toloba.etal.2016,Muller.etal.2019,Mutlu-Pakdil.etal.2022,Mutlu-Pakdil.etal.2024}, while more recent surveys have assembled statistical samples of satellites around MW-like hosts. The Satellites Around Galactic Analogs survey (SAGA; \citealt{Geha.etal.2017}; \citealt{Mao.etal.2024}) uses spectroscopy to identify satellites around 101 MW-like hosts at $25-40$ Mpc. The Exploration of Local VolumE Satellites survey (ELVES; \citealt{Carlsten.etal.2022}) systematically characterizes satellites of $30$ MW-like hosts within $12$ Mpc using deep imaging, yielding $338$ confirmed satellites, reaching a lower mass limit than SAGA.

These efforts have given key insights into the properties of satellites in MW-like environments and helped place the MW within a broader context~\citep{Danieli2023,Greene.etal.2023, Geha.etal.2024, Mao.etal.2024}. However, these MW analogs are not the only environments that host satellites. $\Lambda$CDM predicts that smaller galaxies, even dwarfs that are 1--3 orders of magnitude less massive than the MW, can have their own satellite galaxies~\citep{Sales.etal.2013,Dooley2017,Santos2022}. These lower-mass hosts probe shallower potential wells with less gas in the circumgalactic medium, so the satellite abundance and quenching in these systems test the halo-mass dependence of both subhalo populations and environmental processing. Several surveys have targeted satellites around dwarf and sub-MW-mass hosts, including the Magellanic Analog Dwarf Companions And Stellar Halos survey (MADCASH; \citealt{Carlin2016,Carlin2019,Carlin2025}), the DECam Local Volume Exploration survey (DELVE; \citealt{Drlica2021,McNanna2024,Doliva2025}), and the Large Binocular Telescope Satellites Of Nearby Galaxies Survey (LBT-SONG; \citealt{Garling2021,Davis2021,Davis2024}); see \citealt{Li2025} for a review. These efforts have demonstrated that lower-mass galaxies can host their own satellite populations, but existing samples remain relatively small and heterogeneous in host mass and environment. Consequently, we still lack a large, uniformly selected sample with which to characterize how satellite populations vary with host mass below the MW-mass scale.

Recently, the Exploration of Local VolumE Satellites of Dwarf Galaxies (ELVES-Dwarf; \citealt{Li2025}) and Identifying Dwarfs of MC Analog GalaxiEs~(ID-MAGE; \citealt{Hunter2025}) surveys started to build statistical samples of satellites around dwarf galaxies, with more such systems being identified in other surveys~\citep{Paudel2026}. The ELVES-Dwarf survey~\citep{Li2025,Li2026a,Li2026b} is a systematic census of satellites around dwarf galaxies in the Local Volume. \citet{Li2025} presented the first eight isolated hosts, while the full sample in \citet{Li2026b} comprises 39 hosts with $10^{7} < M_{\star}^{\rm host} < 10^{10}\,M_\odot$ and detects satellites down to $M_\star < 10^{5.7}\,M_\odot$. Complementary work has also emerged from H~\textsc{i}-selected samples. \citet{Zhu2025} identified gas-bearing satellites around late-type hosts spanning dwarf to MW masses and found that their abundance increases with host mass, providing a gas-selected counterpart to optical censuses of satellites.

While these surveys represent a great step toward studying satellites of low-mass hosts, the available sample of hosts remains small. A complementary approach is to measure satellite populations statistically by subtracting foreground and background galaxy counts from large photometric samples \citep[e.g.,][]{Nierenberg.etal.2011,Nierenberg2012,Roberts2021,Wang.etal.2021,Li.etal.2023a,Bhattacharyya.etal.2024,Wang.etal.2025,Pan2026}. This approach enables studies of much larger host samples, at the cost of determining satellite membership for individual systems.

The Merian Survey provides a powerful dataset for this statistical approach, combining deep, wide-field optical imaging over $750\,\mathrm{deg}^2$ with medium-band photometry that enables precise photometric redshift measurements \citep{Luo.etal.2023,Danieli2025}. Merian uses two medium-band filters with the Blanco/Dark Energy Camera (DECam; \citealt{Falugher2015}) to isolate H$\alpha$ and [O~\textsc{iii}] emission at $0.06<z<0.1$. The resulting precise photometric redshifts enable efficient selection of star-forming dwarf galaxies in this redshift range from imaging alone~(Y. Luo et al. 2026, in prep.), while the large survey area provides hundreds of potential hosts for statistical satellite studies. Using Merian, \citet{Pan2026} characterized the satellite populations of $\mathord{\sim}400$ MW analogs with $10^{10.5}<M_{\star,\mathrm{host}}<10^{10.9}\,M_\odot$. In this work, we extend the Merian satellite census to sub-MW-mass hosts ($10^{9.5}<M_{\star,\mathrm{host}}<10^{10.5}\,M_\odot$), filling the previously sparsely sampled regime between dwarf hosts targeted by surveys such as ELVES-Dwarf and ID-MAGE and the better-studied MW-mass population.

\begin{figure*}[t!]
\centering
\includegraphics[width=0.33\textwidth]{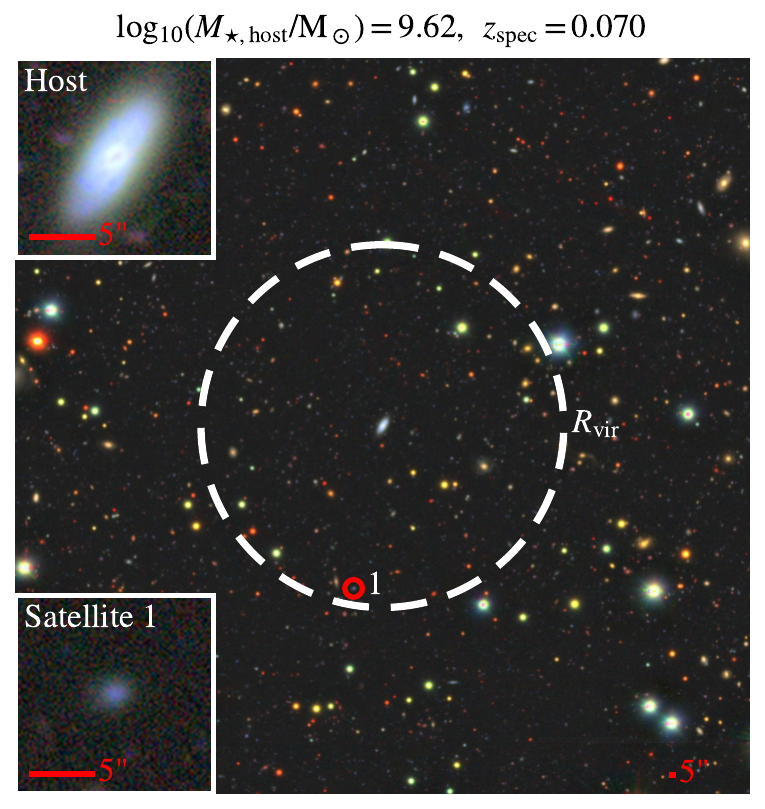}\includegraphics[width=0.33\textwidth]{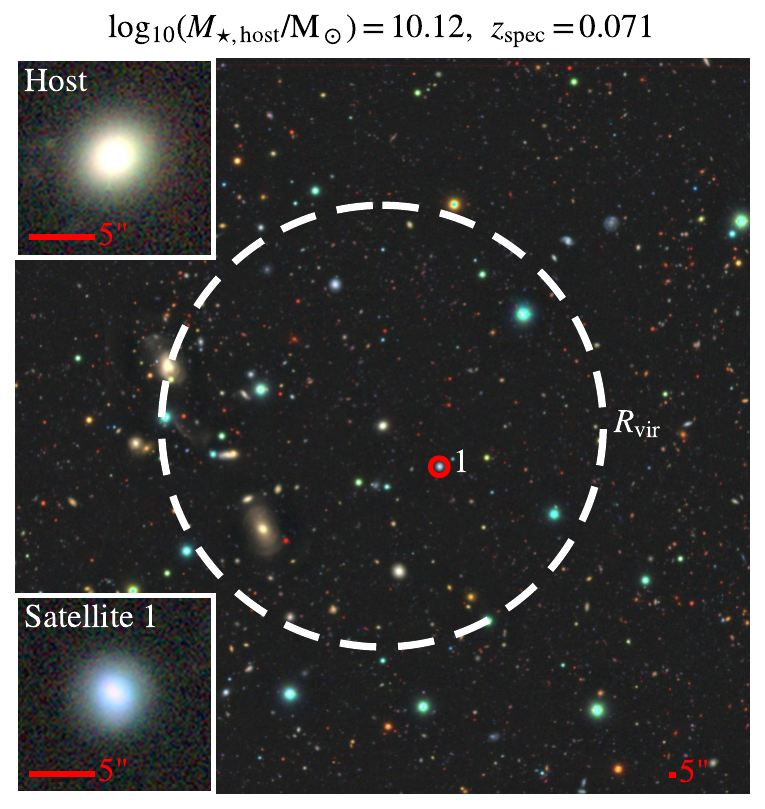}\includegraphics[width=0.33\textwidth]{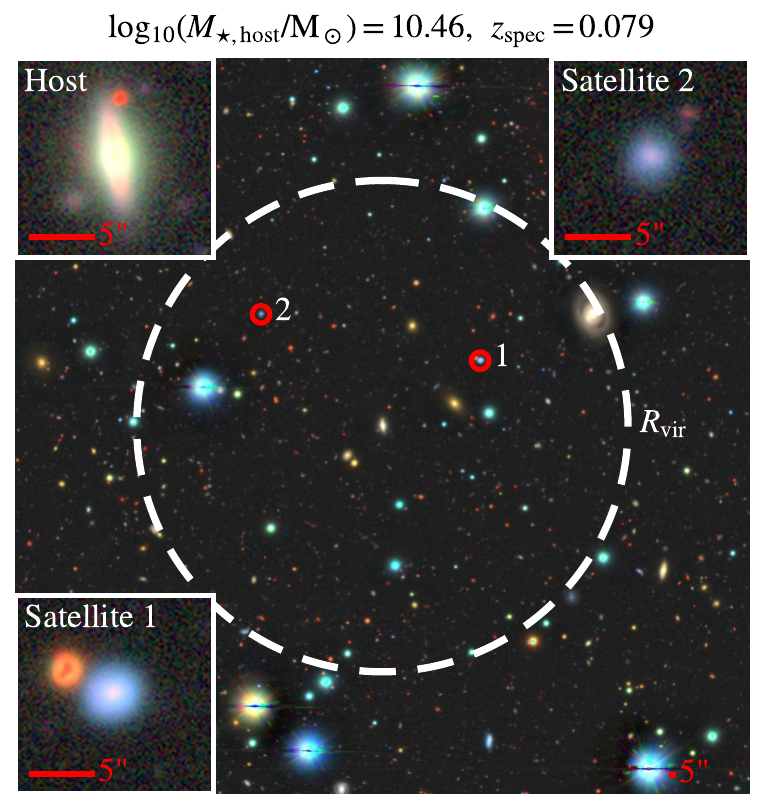}
\caption{HSC color stamps for three representative systems from the Merian sub-MW-mass sample. Hosts span most of the mass range for the sample and contain at least one associated satellite candidate each. Each panel shows the host environment, enclosed by circles representing the virial area, and the position of the host and satellite candidates. The estimated virial radii for the three hosts are $166 \ \rm kpc$, $204 \ \rm kpc$, and $248 \ \rm kpc$ respectively. The corner images show zoomed-in views of each host and candidate.
\label{fig:sample}}
\end{figure*}

This paper is structured as follows. Section \ref{subsec:merian} provides a brief overview of the Merian Survey and its dataset. Section \ref{subsec:hosts} describes the selection criteria of the sub-MW-mass host galaxies, and Sections \ref{subsec:satellites} and \ref{subsec:background} outline the identification of their associated satellite galaxies. Section \ref{sec:statistics} presents our key statistical results on the satellite populations. Finally, Section \ref{sec:conclusions} summarizes our conclusions and discusses the implications of the results.

 Throughout this paper, we adopt a flat $\Lambda$CDM model with $H_0=70 \ \text{km s}^{-1} \ \text{Mpc}^{-1}$ and $\Omega_m=0.3$, and a Salpeter IMF.

\section{SAMPLE SELECTION} \label{sec:samples}

\subsection{The Merian Survey} \label{subsec:merian}

Merian is a wide-field optical imaging survey targeting bright dwarf galaxies over large volumes~\citep{Luo.etal.2023, Danieli2025}. The survey utilizes two medium-band filters, N540 and N708~\citep{Luo.etal.2023}, built for DECam to isolate H$\alpha$ and [O~\textsc{iii}] line emission. These filters are tailored to identify lines corresponding to $0.06\lesssim z \lesssim 0.1$ with high photometric redshift accuracy ($\sigma_{\Delta z/(1+z)}\approx 0.01$).

For this work, we use the first data release of the Merian Survey, which covers $\mathord{\sim}234$ deg$^2$ of its full $\mathord{\sim} 750$ deg$^2$ footprint overlapping with the wide layer of the Hyper Suprime-Cam Subaru Strategic Program (HSC-SSP) equatorial fields~\citep{Aihara.etal.2018a,Aihara.etal.2022}. The public HSC-SSP data then provide us with high spatial resolution wide-field imaging in five broad-band filters ($grizy$). As detailed in \citet{Danieli2025} and \citet{Pan2026}, we perform joint photometry between the two DECam medium-band images and the HSC-SSP $grizy$ images. For accurate color measurements, we employ the Gaussian-Aperture-and-PSF (GAaP) technique~\citep{Kuijken.etal.2008}, which performs PSF- and aperture-matched photometry. GAaP is optimized for consistent measurements across different seeing conditions and telescopes, making it suitable for our data from two different surveys and improving the accuracy of the colors used for the photometric redshifts.

\subsection{Host Selection} \label{subsec:hosts}

We first select sub-MW-mass hosts with robust redshift measurements. The hosts are selected from a cross-matched catalog of sources with spectroscopic redshifts in the range $0.07<z<0.09$, described in \citet{Mintz.etal.2024}. The chosen redshift range is narrower than the full Merian coverage to minimize instrumental effects and ensure the accuracy of the measurements. The host catalog combines the Merian DR1 catalog with spectroscopic data from the Sloan Digital Sky Survey~(SDSS; \citealt{Almeida.etal.2023}), the Galaxy and Mass Assembly survey~(GAMA; \citealt{Baldry.etal.2018}), and Merian spectroscopic observations~(Y. Luo et al. 2026, in prep). 

\begin{figure}[t!]
\centering
\includegraphics[width=\columnwidth]{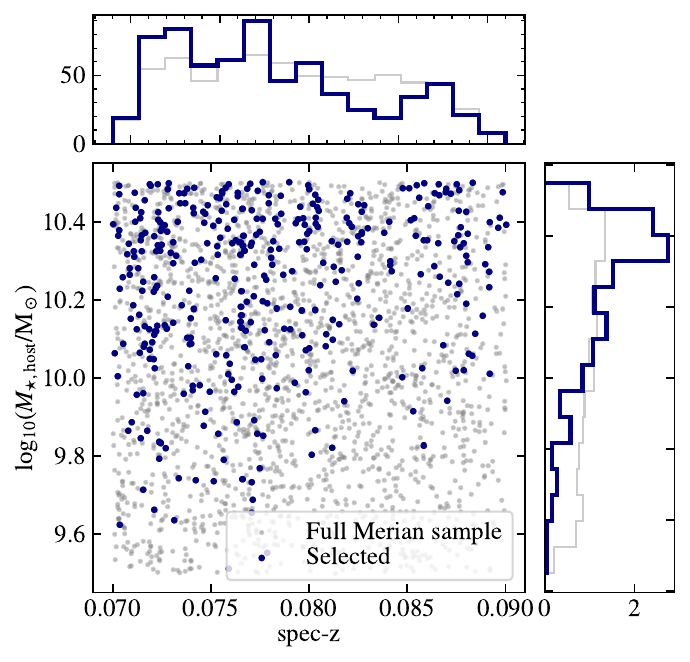}
\caption{Stellar mass and spectroscopic redshift distributions of sub-MW-mass hosts at initial and final steps of sample selection. The gray dots show the initial selection from the Merian DR1 catalog. The blue dots show the final host selection, after cross-matching with SDSS DR13, ensuring they are centrals and are covered by the Merian footprint.
\label{fig:selection}}
\end{figure}

Stellar masses of the hosts are derived from the mass-to-light ratio relation in \citet{Bell.etal.2003} and photometric data from HSC-SSP:
\begin{equation}
    \log(M_\star/L_r) = -0.306 + 1.097(g-r)
\end{equation}

Sub-MW-mass hosts are then identified based on redshift, stellar mass, and group central flags constructed by applying the group finder in \cite{Lim.etal.2017} to Sloan Digital Sky Survey Data Release 13 (SDSS DR13):
\begin{enumerate}
    \item $9.5 < \log[M_{\star,\text{host}}/M_\odot] < 10.5$
    \item $0.07 < z < 0.09$
    \item \texttt{is\_central} in SDSS DR13 using the group finder in \citet{Lim.etal.2017}
\end{enumerate}

This mass range complements the previous studies on Milky Way analogs by \citet{Pan2026} while still falling within the range covered by the group finder. The scarcity of lower-mass hosts in public group catalogs makes it difficult to push this limit further.

The third condition ensures all hosts are central galaxies of their halos, removing contamination from massive satellites of larger groups. An initial number of $2342$ sub-MW-mass hosts are identified in the Merian DR1 catalog after the first two cuts. Out of those, $880$ are cross-matched with SDSS DR13, and $641$ are classified as central in the group finder. Most hosts in the lower end of the stellar mass range were not retained after the cross-matching, with the group catalog being incomplete at these masses. After cross-matching, the \texttt{is\_central} cut mainly affected the higher end. 

Finally, we exclude hosts whose virial area is not fully covered by Merian full-color-full-depth. To do so, we generate a Hierarchical Equal Area isoLatitude Pixelization (\texttt{HEALPix}; \citealt{Gorski.etal.2005}) projection map for Merian DR1. The \texttt{HEALPix} map includes regions with data in all seven bands ($grizyN540N708$; full color) and with effective exposure times $t_{\text{eff}}>1200$ sec for $N708$ and $t_{\text{eff}}>1800$ sec for $N540$ (full depth). The effective exposure time $t_{\text{eff}}$ is a measure of effective depth accounting for observing conditions~\citep{Danieli2025}. This map has a resolution of $\mathord{\sim} 50''$, which is significantly smaller than the Merian DR1 full-color-full-depth field of view of $234 \ \text{deg}^2$. We then use the \texttt{HEALPix} map to calculate the fraction of the total projected virial area of each host that is masked due to artifacts. We exclude any host with more than $10\%$ of its virial area masked. This reduces our total host count to $357$. Figure~\ref{fig:sample} shows HSC color stamps for three sample host-satellite systems. The $M_{\star,\text{host}}$ and photo-$z$ distributions of these hosts are illustrated in Figure~\ref{fig:selection}. The host stellar mass distribution highlights how the sample is incomplete at lower masses, with most hosts falling within the $M_{\star, \rm host}>10^{10}\ \rm M_\odot$ region. Most low-mass hosts identified in the initial Merian sample are absent from the group catalog.

\subsection{Satellite galaxy selection} \label{subsec:satellites}
\subsubsection{Magnitude, Color, and Size Cuts} 

The candidate satellites are selected from the Merian DR1 photometric catalog. For this study, we include every source in the Merian DR1 photometric catalog where the source:
\begin{enumerate}
    \item is not masked by the HSC-SSP PDR3 bright star mask
    \item is in the Merian full-depth full-color area
    \item meets the photometry quality criteria from \citet{Danieli2025}
    \item is not a star according to its $i$-band extendedness value 
\end{enumerate}

First, we estimate host halo masses based on the stellar-to-halo mass relation~(SHMR) from \citet{Rodriguez2017}. We then use \texttt{Halotools}~\citep{Hearin.etal.2017} to compute each halo's virial radius ($R_{\text{vir}}$). This corresponds to the radius that encloses a mean density $\rho(z) = \Delta_{\text{vir}}(z)\rho_c(z)$, where $\rho_c$ is the critical density of the universe, and $\Delta_{\text{vir}}$ is an overdensity factor~\citep{Bryan1998}. The host virial radii range from $158$ to $256$ kpc, with a mean of $219$ kpc, corresponding to angular radii of $1.81’$–$3.16’$ on the sky.

Uncertainties in star formation models can lead to errors in the stellar masses measured from color~\citep{Conroy2009,Roediger2015,delosReyes2025}. The choice of stellar population synthesis models, IMF, and dust law can lead to systematic shifts in the measurements. We estimate the systematic uncertainty associated with the host stellar masses to be of the order $0.2 \ \text{dex}$. Furthermore, there is an upper bound of $0.16 \ \text{dex}$ on the intrinsic scatter of the SHMR~\citep{Tinker2017}. These factors combined lead to an average uncertainty on the true virial radii of $R_{\rm vir, true}/R_{\rm vir}=1.0^{+0.3}_{-0.2}$. Appendix~\ref{sec:rvir} describes the changes to the sample when the search is expanded to the upper limit of this range.

\begin{figure}[t!]
\centering
\includegraphics[width=\columnwidth]{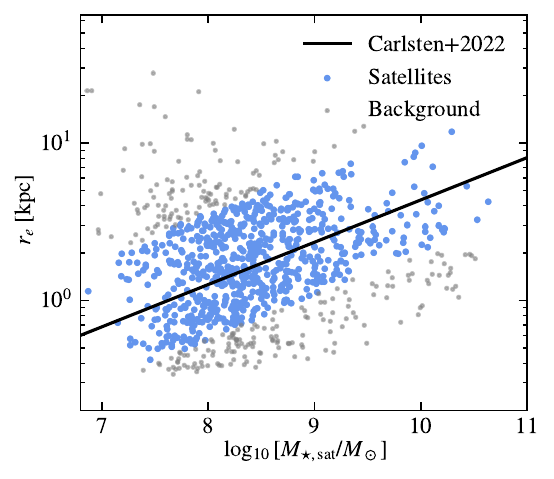}
\caption{Size-mass relations of selected satellite candidates. The solid black line shows the size-mass relation derived by \citet{Carlsten.etal.2021}. The blue dots represent the accepted region, while the gray dots are excluded.
\label{fig:size_mass}}
\end{figure}

We identify an initial sample with a cut of $r < 1.0\,R_{\text{vir}}$. We do not apply an inner radial cut, but any contamination within the inner regions of the host is expected to be removed during visual inspection. We then apply three cuts based on \texttt{cModel} magnitude, \texttt{GAaP} color, and $r_e$ to remove background objects and unresolved sources:
\begin{equation}
    i_{\text{cModel}}<23 \ \text{mag}, \ 0<(g-r)_{\text{GAaP}}<1, \ r_{e}>0.25'',
\end{equation}
where $r_e$ is the half-light radius determined from the quadrupole moments of the shape parameters, derived from the $N708$-band. \texttt{CModel} photometry~\citep{Bosch.etal.2018} estimates the total flux by fitting the object with a linear combination of exponential and de Vaucouleurs surface-brightness profiles. The magnitude cut ensures that we only include relatively bright sources, minimizing unresolved background objects and avoiding failures in the photo-$z$ data. 

The color cut selects sources with plausible colors for low-mass galaxies, excluding extremely red or blue objects~\citep{Greco.etal.2018}.

The angular size cut $r_e > 0.25''$ excludes unresolved sources that are likely background galaxies. We aim to include satellites down to masses of $\log[M_{\star,\rm sat}/M_\odot]=7.5$, where the Merian photo-$z$ catalog is expected to be $\mathord{\sim} 50\%$ complete~(Y. Luo et al. 2026, in prep.), as derived from the COSMOS 30-band catalog~\citep{Ilbert2009, Weaver2022}. Using the average mass--size relation of satellites from \citet{Carlsten.etal.2021}, a typical satellite galaxy of $\log[M_{\star,\rm sat}/M_\odot]=7.5$ has a physical size of $\mathord{\sim} 0.8$ kpc, corresponding to an angular size of $\mathord{\sim} 0.5''$ at $z=0.08$. The size cut $r_e > 0.25''$ then ensures such satellites are kept in our sample, including those $1\sigma$ below the average trend. This corresponds to a looser size cut than \citet{Pan2026} ($r_e > 0.5''$), aiming to accept lower-mass sources for the sub-MW-mass hosts. Any contamination from this looser cut is expected to be eliminated by the visual inspection and statistical background suppression processes in Section~\ref{subsec:background}.

\subsubsection{Photometric Redshift Cut}

We adopt the Merian DR1 photo-$z$ catalog (Y. Luo et al. 2026, in prep) for all of our sources. Photometric redshifts in the Merian DR1 photo-$z$ catalog are derived with the template-fitting photo-$z$ code \texttt{EAZY}~\citep{Brammer.etal.2008}. The \texttt{EAZY} output includes the $\chi^2$ distribution of photo-$z$ across 184 points in the $0 < z < 1.5$ range. We use this distribution to derive the probability density function~(PDF) of the photo-$z$, $p(z)$.

We then integrate $p(z)$ over the in-band Merian range of $0.06 < z < 0.1$. We only include sources with integrated $p(z) > 0.26$. This threshold is expected to provide an optimal balance of purity and completeness, with an estimated $\mathord{\sim} 75\%$ of sources lying within the in-band region\citep{Pan2026}. 

\subsubsection{Size--Mass Cut}

To further exclude foreground and background sources, we apply a cut based on the size--mass relation for late-type dwarf galaxies derived from the Exploration of Local VolumE Satellites (ELVES) Survey:
\begin{equation}
    \log(r_e/\text{pc}) = a + b\log(M_\star/M_\odot)
\end{equation}
with $a=0.805^{+0.242}_{-0.245}, \ b=0.282^{+0.034}_{-0.033}$~\citep{Carlsten.etal.2021}. We take the relation for late-type galaxies since Merian probes star-forming systems. We convert angular sizes of the sources into physical sizes by assuming the same redshift as the host.

The stellar masses of dwarf satellites are estimated with the same color-$M_\star/L_r$ relation as the hosts~\citep{Bell.etal.2003}, which was found to be the most accurate for our mass range~\citep{de.los.Reyes.etal.2024}. We adopt a flat uncertainty of $\sigma = 0.1 \ \text{dex}$ in the size-mass relation and exclude sources that deviate by more than $5\sigma$ above or $3\sigma$ below the line. The range of larger objects accepted is wider to account for ultradiffuse galaxies~\citep{van.Dokkum.etal.2015,Greco.etal.2018,Danieli.etal.2019,Danieli.etal.2022,Li.etal.2023a,Li.etal.2023b}. Given the assumed relation and uncertainty, less than $0.2\%$ of real satellites should be excluded by these cuts. The results of this cut are shown in Figure \ref{fig:size_mass}. The figure also shows that a small fraction of the satellites lie below the $\log[M_{\star,\rm sat}/M_\odot]=7.5$ threshold. To avoid the region where the photo-$z$ catalog is expected to be less than $50\%$ complete, we exclude any such low-mass sources from the sample to avoid regions with low completeness.

Any remaining satellite candidates that have been assigned to more than one host are only considered to be associated with the closest host. Table~\ref{tab:cuts} summarizes the number of sources remaining after each cut described in this section. After applying all cuts, the number of satellite candidates remaining is $654$.

\subsection{Deblending and Background correction}\label{subsec:background}

Before finalizing the sample selection, five authors in the Merian collaboration visually inspected the $654$ galaxies based on their HSC-SSP $gri$ images. Contaminants such as shredded fragments of massive objects, blended objects, artifacts, or sources affected by bright nearby objects were identified and excluded. Furthermore, we remove any satellite candidates with greater stellar mass than their host. After removing these objects, we obtain a final sample of $471$ sources as satellite candidates for the 357 sub-MW-mass hosts. This sample is used for all the following analyses.

\begin{deluxetable}{cc}[t!]
\tablehead{
\colhead{\hspace{1.2cm}Selection step\hspace{1.2cm}} & \colhead{\hspace{1.2cm}Sources remaining\hspace{1.2cm}}
}
\startdata
Initial sample ($d_{\rm proj}<R_{\rm vir}$)                 &   51,665 \\
Color, magnitude, size cuts                &   32,083 \\
Integrated $p(z)$ cut     &  976 \\
Size--mass cut                   &    702\\
Duplicates            &    687 \\
$M_{\star,\rm sat}>10^{7.5} \ \rm M_\odot$            &    654 \\
Deblending                   &    471 \\
Background subtraction                   &    148 $\pm$7 \\
\enddata
\caption{Number of sources remaining after each step in the satellite selection process.\label{tab:cuts}\vspace{-0.7cm}}
\end{deluxetable}

While the Merian photometric redshifts efficiently select galaxies within the target redshift range, their line-of-sight distance uncertainties are much larger than the physical scale of individual host halos. Consequently, some satellite candidates are inevitably foreground or background galaxies projected within the host virial radius. \citet{Pan2026} estimated the global background number density of the Merian sources to be $\mathord{\sim} 88/\rm deg^2$ by selecting $60$ random circles on the sky and applying the same photometric cuts and visual inspection procedure used for satellite selection. However, this global averaging of background contributions introduces systematic biases, as hierarchical structure formation leads to some regions having significantly higher densities than others~\citep{White1978}. We thus account for these variations with a local estimate of background contamination~\citep{Chen2006,Wang2014,Alpaslan2020,Tinker.etal.2021,Li.etal.2023a,Bhattacharyya.etal.2024,Cataldi2026}. 

For each host, we select sources within an annular region of $2.5<r/R_{200,\rm c} < 4.5$, where $R_{200,\rm c}$ represents the estimated radius within which the mean enclosed density is $200$ times the critical density. The lower limit is chosen to be the estimated border for the region with backsplash and infalling satellites~\citep{Buck2019,Diemer2021} to avoid including associated galaxies. The high upper limit ensures that the background is sampled from a large area (approximately $7.7$ times larger than the virial area), minimizing Poisson error. We then apply the satellite selection process described in Section \ref{subsec:satellites} for each annular region, assuming the same photo-$z$ value as the host for each source.

This procedure identifies a total of $2766$ sources, and an additional $318$ are removed after visual inspection, giving us a total of $2448$ background sources across the $357$ hosts. For a given host, we divide the number of identified background sources by the area of the corresponding annular region to obtain the local background number density, and multiply it by the virial area to obtain the final contamination. The median number density is approximately $144/\rm deg^2$, which is 1.6$\times$ larger than the result from \citet{Pan2026}. This difference can partially be attributed to the stricter size cut of $r_e>0.5''$ in the previous study, but the main effects are expected to be the biases introduced by global averaging and low statistics. We then subtract the local background estimate from each host individually. This local quantification of the background should account for physical differences in matter distribution across the observed region and provide a more accurate estimate of the satellite population. 

Adding the estimated contamination from all hosts and assuming Poisson error for the initial identification\footnote{We model the number of background sources identified in each annulus as an independent Poisson random variable. The large area of each annulus leads to reduced Poisson error~(median of $\mathord{\sim} 40\%$ per host), and subsequently to a low uncertainty in the total background subtracted.}, we find a total background number of $323\pm7$, bringing the total number of satellite candidates in our sample after background subtraction to $148\pm 7$.

For a more direct comparison with the sample from \citet{Pan2026}, we repeat the local background estimation from Section~\ref{subsec:background} around the $393$ MW-mass hosts. We identify $6536$ sources in these annuli and assume a $\mathord{\sim} 27\%$ contamination rate based on the visual inspection conducted by \citet{Pan2026}. Removing the estimated local background from each individual host reduces the total number of satellites of the MW sample to $161\pm 8$. The total background estimated with this procedure represents a $\mathord{\sim} 85\%$ increase over the original global estimate. This increase can largely be attributed to an overall shift in the background distribution due to the hosts generally existing in high-density environments. Furthermore, this local approach takes into account the prominent high-end tail in background, which is not represented by the median density.

\section{RESULTS}\label{sec:statistics}

\begin{figure}[t!]
\centering
\includegraphics[width=\columnwidth]{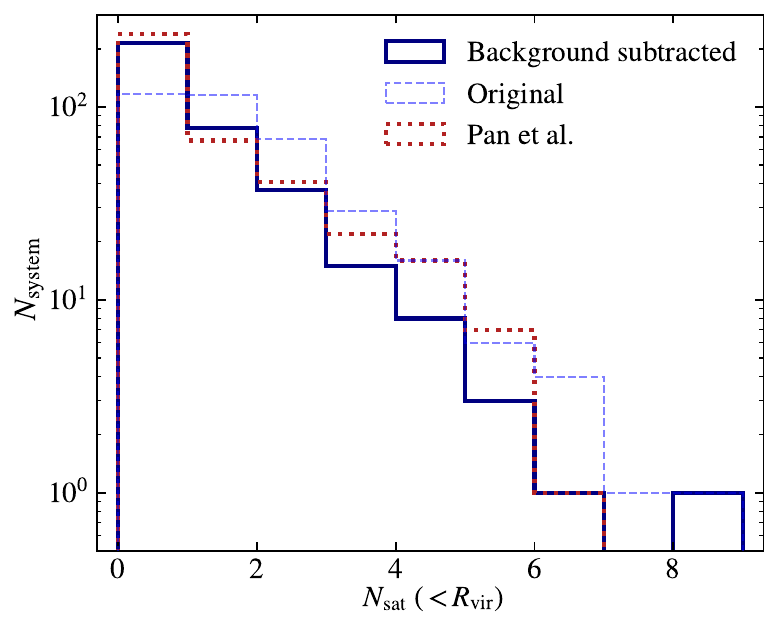}
\caption{Distribution of number of satellites across all hosts. The blue dashed lines represent the numbers of satellites after the cuts from Section \ref{subsec:satellites}. The solid navy lines have the background satellites identified in Section \ref{subsec:background} subtracted. The red dotted lines represent the results from \citet{Pan2026} with a local background estimate subtracted.
\label{fig:Nsat_Nsys}}
\end{figure}

\subsection{Satellite abundance}\label{subsec:abundance}

\begin{figure*}[t!]
\centering
\includegraphics[width=0.33\textwidth]{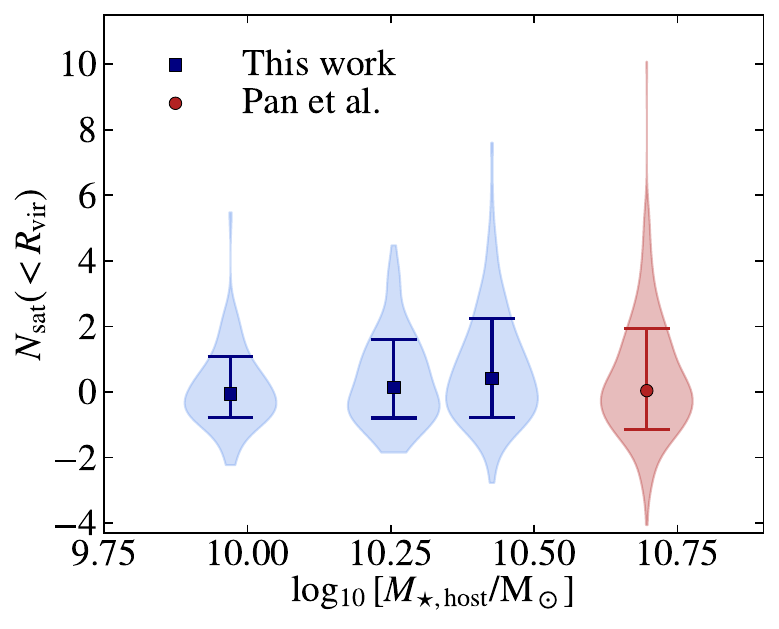}\includegraphics[width=0.33\textwidth]{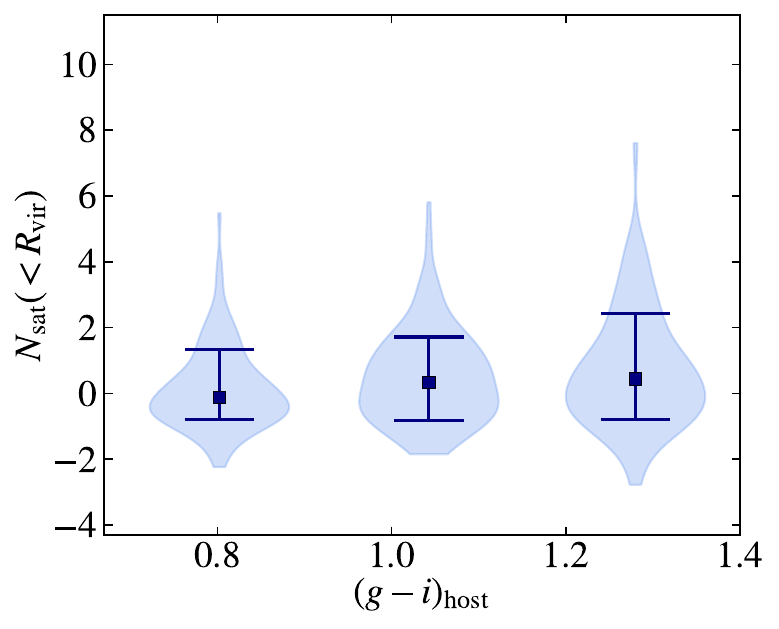}\\\includegraphics[width=0.33\textwidth]{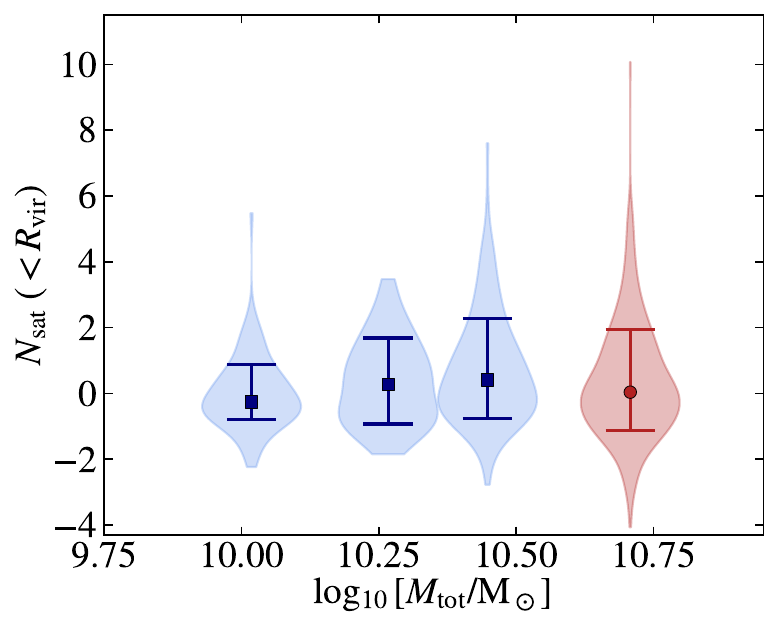}\includegraphics[width=0.33\textwidth]{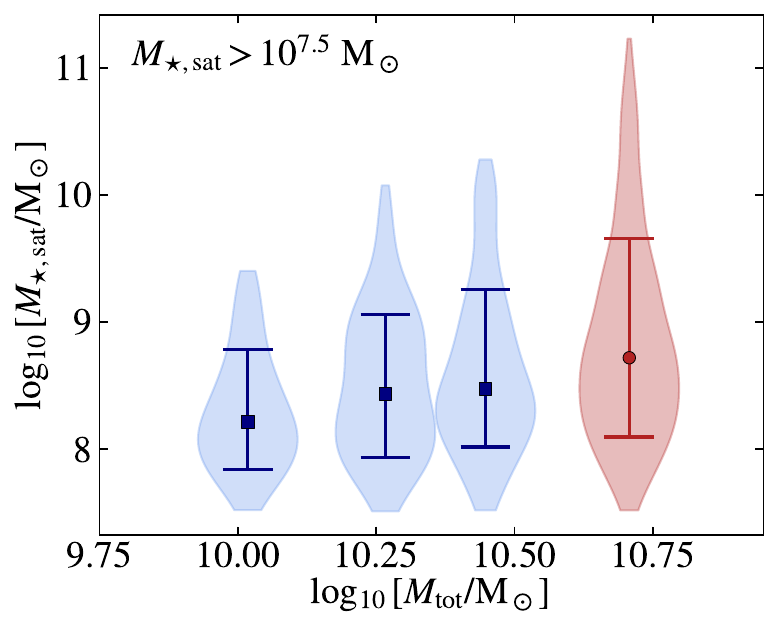}
\caption{\textbf{Top left:} Satellite abundance $N_{\text{sat}}$ as a function of host stellar mass $M_{\star,\rm host}$. The blue bins correspond to: $M_\star < 10^{10.1} \ \rm M_\odot$, $10^{10.1} \ \text{M}_\odot \leq M_{\star, \rm host} < 10^{10.35} \ \rm M_\odot$, and $M_{\star, \rm host} \geq 10^{10.35} \ \rm M_\odot$. The median $N_{\text{sat}}$ in each bin is shown as a square with the $16^{\rm th}$ and $84^{\rm th}$ percentiles as horizontal bars. \textbf{Top right:} $N_{\text{sat}}$ as a function of host color $(g-i)_{\rm host}$. The blue bins correspond to: $(g-i)_{\rm host} < 0.9$, $0.9 \leq (g-i)_{\rm host} < 1.2$, and $(g-i)_{\rm host} \geq 1.2$. \textbf{Bottom left:} $N_{\text{sat}}$ as a function of system total stellar mass $M_{\text{tot}}$. The blue bins correspond to: $M_{\rm tot} < 10^{10.15} \ \rm M_\odot$, $10^{10.15} \ \text{M}_\odot \leq M_{\rm tot} < 10^{10.35} \ \rm M_\odot$, and $M_{\rm tot} \geq 10^{10.35} \ \rm M_\odot$. \textbf{Bottom right:} satellite stellar mass $M_{\star,\text{sat}}$ as a function of $M_{\text{tot}}$, split into the same bins. For all distributions, the Merian sub-MW-mass sample shows some correlation between the median and $84^{\rm th}$ percentiles and the given property. The single red bin corresponds to the data for MW-mass hosts by \citet{Pan2026}. Only satellites with $M_{\star}^{\rm sat}>10^{7.5} \ \rm M_\odot$ are included.
\label{fig:violins}}
\end{figure*}

This section explores how the number and spatial distribution of satellites in our sample are affected by the host properties. Recent observations have revealed rich groups with more satellites than expected given theoretical predictions in $\Lambda$CDM, from the dwarf scale~\citep{Li2026a} to MW-mass~\citep{Homma2024,Muller2024,Kanehisa2024,Taibi2025}. However, the region between these two mass scales remains sparsely explored. We thus investigate the number of satellites identified in our hosts that fall in the defined sub-MW-mass range.

Figure \ref{fig:Nsat_Nsys} illustrates how the satellite abundance $N_{\text{sat}}$, the number of satellites assigned to a host, is distributed across our $357$ hosts. We estimate the background-subtracted satellite abundance $N_{\text{sat}}$ while accounting for the Poisson noise in the low-count regime. Following the procedure by \citet{Pan2026} (see e.g. their Appendix B), we model the posterior distribution of the true satellite counts $\lambda$ as a Gamma distribution with shape parameter $k_{\text{obs}}-b+0.5$, where $k_{\text{obs}}$ is the number of satellite candidates identified in our sample and $b$ is the estimated background count. Taking $N_{\text{sat}}$ to be the mean of the distribution, we find:

\begin{equation}
    N_{\text{sat}} = k_{\text{obs}}-b+0.5.
\end{equation}

This $N_{\text{sat}}$ distribution shows a broad range of satellite populations around the Merian hosts. After background subtraction, $\mathord{\sim} 60\%$ of hosts have $N_{\rm sat}< 1$, $\mathord{\sim} 22\%$ have $1\leq N_{\rm sat}< 2$, $\mathord{\sim} 10\%$ have $2\leq N_{\rm sat}< 3$, and $\mathord{\sim} 8\%$ have $N_{\rm sat} > 3$. Figure~\ref{fig:Nsat_Nsys_13R} illustrates how the distribution shifts to the right when the search radius is expanded to $1.3R_{\rm vir}$.

The resulting background-subtracted $N_{\text{sat}}$ distribution is overlaid in Figure~\ref{fig:Nsat_Nsys}. The MW-mass hosts show a significantly more prominent tail, with the distribution consistently above the sub-MW sample for high values of $N_{\text{sat}}$. This is also reflected by the fact that a larger fraction of hosts ($\mathord{\sim} 12\%$) have more than $3$ satellites.

Galaxies with higher stellar mass generally reside in more massive halos, which can host a greater number of satellites~\citep{Behroozi.etal.2010,Moster.etal.2013}. To explore how host stellar mass correlates with satellite abundance in the Merian sub-MW-mass sample, the top left panel of Figure~\ref{fig:violins} shows the connection between $N_{\text{sat}}$ and host stellar mass $M_{\star,\text{host}}$. The hosts are divided into three bins: $M_{\star, \rm host} < 10^{10.1} \ \rm M_\odot$, $10^{10.1} \ \text{M}_\odot \leq M_{\star, \rm host} < 10^{10.35} \ \rm M_\odot$, and $M_{\star, \rm host} \geq 10^{10.35} \ \rm M_\odot$. The figure shows correlation between median $N_{\text{sat}}$ and $M_{\star, \rm host}$, with the three bins having median and $68\%$ containment values of $-0.1^{+1.2}_{-0.7}$, $0.1^{+1.5}_{-0.9}$, and $0.4^{+1.8}_{-1.2}$. The $84^{\rm th}$ percentiles also show correlation with $M_{\star, \rm host}$, with more systems reaching higher $N_{\rm sat}$ at higher masses. The Pearson correlation coefficient for the medians is $\rho=0.97$. The $16^{\rm th}$ percentile in each bin is largely determined by the background, with higher-mass systems often having greater numbers subtracted due to their larger virial area. The left panel of Figure~\ref{fig:violins_13R} illustrates how these trends are still present when the search radius for satellites is expanded to $1.3R_{\rm vir}$. The distribution for the MW-mass hosts found in Merian with the local background estimation is shown in red~\citep{Pan2026}, with a range of $0.0^{+2.0}_{-1.1}$. Including this median value decreases the correlation to $\rho=0.31$. This slight decrease in median and $84^{\rm th}$ percentile relative to the sub-MW-mass sample can largely be attributed to the stricter size cut of $r_e>0.5''$ in \citet{Pan2026}. The red bin's high-$N_{\rm sat}$ tail is still the most prominent, as reflected in Figure~\ref{fig:Nsat_Nsys}.

A similar trend is shown in the top right panel of Figure~\ref{fig:violins}, which illustrates the connection between $N_{\text{sat}}$ and host color. The hosts are divided into three bins:  $(g-i)_{\rm host}$. $(g-i)_{\rm host} < 0.9$, $0.9 \leq (g-i)_{\rm host} < 1.2$, and $(g-i)_{\rm host} \geq 1.2$. The two properties are also correlated, with the median $68$ percentile ranges for the three bins being $-0.1^{+1.2}_{-0.9}$, $0.3^{+1.7}_{-1.1}$, and $0.4^{+2.0}_{-1.2}$. The correlation coefficient for the median values is $\rho=0.998$. The middle panel of Figure~\ref{fig:violins_13R} illustrates how these trends also hold for $N_{\rm sat}(<1.3R_{\rm vir})$.

\begin{figure*}[t!]
\centering
\includegraphics[width=\columnwidth]{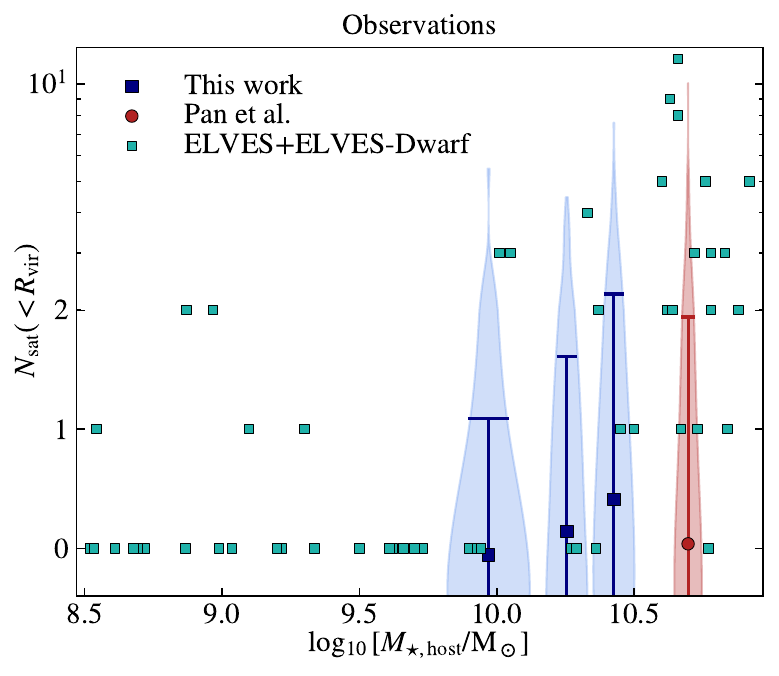}\includegraphics[width=\columnwidth]{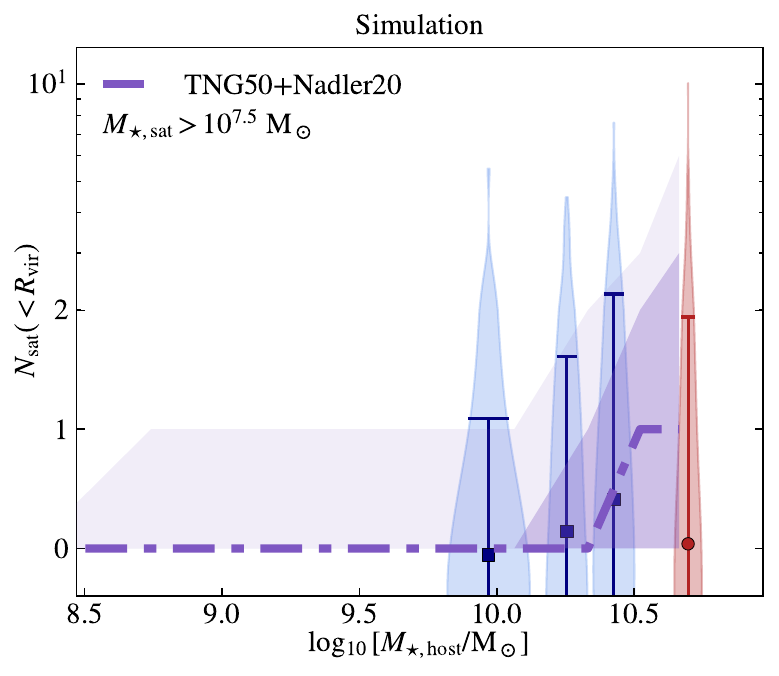}
\caption{Satellite abundance $N_{\text{sat}}$ as a function of host stellar mass $M_{\star,\text{host}}$. \textbf{Left:} The $N_{\text{sat}}$ distributions for the Merian sub-MW-mass sample is shown in three blue host mass bins: $M_{\star,\rm host} < 10^{10.1} \ \rm M_\odot$, $10^{10.1} \ \text{M}_\odot \leq M_{\star, \rm host} < 10^{10.35} \ \rm M_\odot$, and $M_{\star, \rm host} \geq 10^{10.35} \ \rm M_\odot$. The median $N_{\text{sat}}$ in each bin is shown as a square, with the $16^{\rm th}$ and $84^{\rm th}$ percentiles of $N_{\text{sat}}$ shown as horizontal bars. Satellite abundance seems to be correlated with host stellar mass. We include the number of confirmed satellites with $M_\star>10^{7.5} \ \text{M}_\odot$ for hosts in ELVES and ELVES-Dwarf surveys, shown as turquoise squares. We also include the results from \citet{Pan2026} as red points and bars. \textbf{Right:} Theoretical predictions of $N_{\rm sat}$ from TNG50 using the SHMR from \citet{Nadler2020}. The dashed-dotted line represents the median, with the dark and faint bands showing the $1\sigma$ and $2\sigma$ ranges, respectively. The medians of the Merian bins fall close to or below the TNG50 predictions, but the $84^{\rm th}$ percentiles are universally higher in the observations, indicating a possible overabundance of rich groups in the sample.
\label{fig:Mstar}}
\end{figure*}

To further explore the connection between environment and satellite properties, the bottom left panel of Figure \ref{fig:violins} shows the connection between $N_{\text{sat}}$ and system total stellar mass $M_{\text{tot}}$. The total stellar mass of the system is the sum of the host's stellar mass and the stellar mass of all of the associated satellites. This total mass has been shown to be correlated with group mass~\citep{Carlsten.etal.2022,Mao.etal.2024,Pan2026}. This quantity will often be overestimated in the statistical sample, as satellite candidates may be part of the background. To correct for this effect, we scale the sum of the satellite masses by a fixed fraction $\langle \frac{N_{\rm sat}}{k_{\rm obs}}\rangle$, representing the fraction of candidates expected to be actual satellites, averaged across all hosts with $k_{\rm obs}>0$. This value corresponds to $\mathord{\sim} 74\%$ for the sub-MW-mass sample. The resulting total mass is thus defined as:

\begin{equation}
    M_{\rm tot} = M_{\star, \rm host} + \Big\langle \frac{N_{\rm sat}}{k_{\rm obs}}\Big\rangle\sum_i M_{\star, \text{sat}_i}
\end{equation}

In the figure, the hosts are divided into three bins: $M_{\rm tot} < 10^{10.15} \ \rm M_\odot$,  $10^{10.15} \ \rm M_\odot \leq M_{\rm tot} < 10^{10.35} \ \rm M_\odot$, and $M_{\rm tot} \geq 10^{10.35} \ \rm M_\odot$. The plot shows a correlation between both median and $84^{\rm th}$ percentile of $N_{\text{sat}}$ and $M_{\rm tot}$, with the median and $68$ percentile interval for the three bins being $-0.3^{+1.1}_{-0.5}$, $0.3^{+1.4}_{-1.2}$, and $0.4^{+1.9}_{-1.2}$. The correlation coefficient for the medians is $\rho=0.97$. The right panel of Figure~\ref{fig:violins_13R} illustrates how these trends also hold for $N_{\rm sat}(<1.3R_{\rm vir})$. The trend does not extend to the MW-mass hosts, with the corresponding values being $0.0^{+2.0}_{-1.1}$. Including this median value decreases the correlation to $\rho=0.40$. However, the higher-mass bins consistently have wider tails at higher values of $N_{\rm sat}$. 

Finally, the bottom right panel shows some correlation between median $\text{log}_{10}(M_{\star, \text{sat}}/\rm M_\odot)$ and $M_{\rm tot}$, with the median and $68\%$ containment values for the three bins being $8.2^{+0.6}_{-0.4}$, $8.4^{+0.7}_{-0.5}$, and $8.5^{+0.8}_{-0.5}$. The correlation coefficient for the medians is $\rho=0.96$. The corresponding value for the MW-mass hosts is $8.7^{+0.9}_{-0.7}$, which extends the trend, increasing the correlation coefficient to $\rho=0.99$. The higher-mass bins again show an enhancement in their tails at high values of $M_{\star, \text{sat}}$, which is reflected by the $84^{\rm th}$ percentiles. The $M_{\star, \text{sat}}$ distributions are limited by the low completeness below $10^8 \text{M}_\odot$.

Figure~\ref{fig:Mstar} compares the satellite abundances from this work with data from other surveys and theoretical predictions. All ELVES systems with coverage radius $>0.9\,R_{\rm vir}$ are overlaid with the bins~\citep{Carlsten.etal.2022}, shown as turquoise squares, indicating the number of satellites with $M_{\star,\rm sat}>10^{7.5} \ \rm M_\odot$. These systems show a correlation between $N_{\rm sat}$ and $M_{\star,\rm host}$ and populate similar regions to the Merian distributions, but with many rich groups. Unlike ELVES, the Merian samples have significantly larger uncertainties for quenched satellites, which may make up $\mathord{\sim} 70\%$ of sources at $M_{\star, \rm sat}=10^{7.5} \ \rm M_\odot$~\citep{Greene.etal.2023}. This effect is expected to shift the ELVES sample up relative to Merian. Systems from ELVES-Dwarf are also shown~\citep{Li2025,Li2026b}, indicating an overall decline in the number of satellites with $M_{\star,\rm sat}>10^{7.5} \ \rm M_\odot$ at lower host masses, though a few rich groups still exist.

\begin{figure*}[t!]
\centering
\includegraphics[width=\columnwidth]{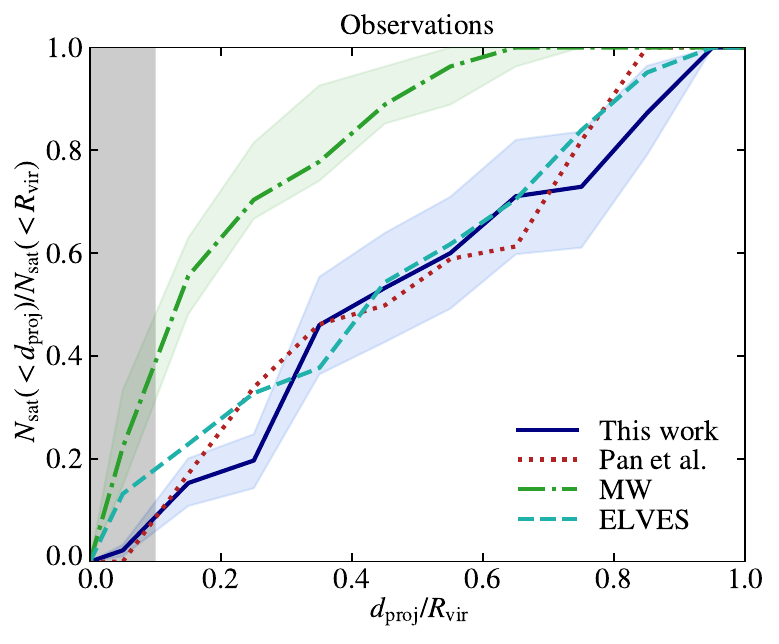}\includegraphics[width=\columnwidth]{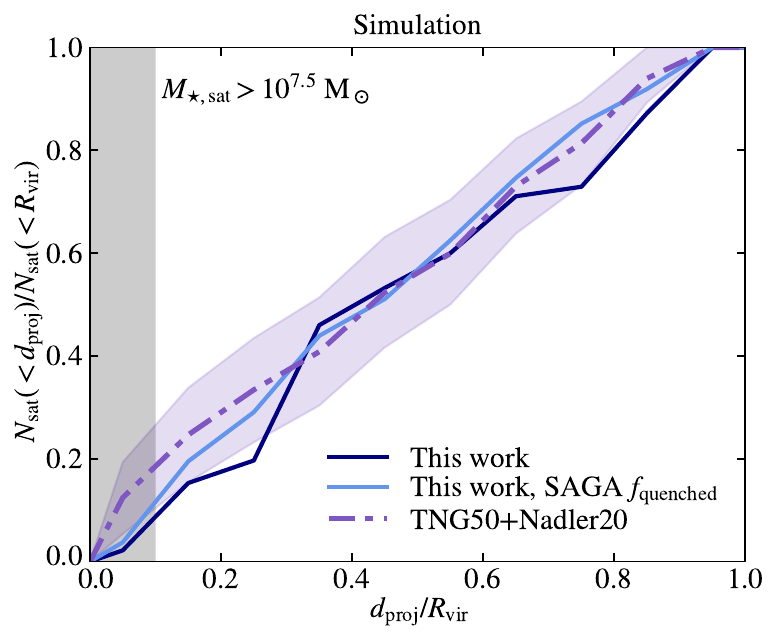}
\caption{Cumulative radial distribution of satellites. \textbf{Left}: average distribution for the Merian background-subtracted sub-MW-mass sample~(navy solid line), MW~(green dash-dotted line), ELVES confirmed satellites~(turquoise dashed line, $M_{\star, \rm sat}\geq 10^{7.5} \ \rm M_\odot$), and Merian background-subtracted, completeness-corrected MW-mass sample (red dotted line, \citealt{Pan2026}). The blue and green bands represent the $1\sigma$ spread of the Merian sample and MW respectively. The sub-MW-mass sample is similar to the MW-mass and ELVES samples. \textbf{Right:} Merian background-subtracted sub-MW-mass sample (navy solid line), Merian background-subtracted, completeness-corrected sub-MW-mass sample (blue solid line), and TNG50 predictions with SHMR from \citet{Nadler2020} (purple dash-dotted line, $M_{\star, \rm sat}\geq 10^{7.5} \ \rm M_\odot$). The purple band corresponds to the $1\sigma$ range of the TNG50 predictions. The Merian sub-MW-mass sample largely agrees with TNG50 predictions, and completeness correction leads to a slight central enhancement. The sub-MW-mass Merian sample is expected to be significantly incomplete for $d_{\rm proj} < 0.1R_{\rm vir}$, and the MW-mass sample excludes satellites in this region entirely.
\label{fig:Rhost}}
\end{figure*}

The right panel of Figure~\ref{fig:Mstar} shows the theoretical predictions of $N_{\rm sat}$ from the TNG50 cosmological simulation~\citep{Nelson2019}. The data are drawn from the TNG50-1 hydrodynamic run, which simulates structure formation in a $50 \ \rm Mpc$ box with dark matter mass resolution of $m_{\rm DM}=4.5\times 10^5 \ \rm M_\odot$, which is enough to resolve the subhalos found in the Merian sample. Halos are identified with the friends-of-friends algorithms and \textsc{SubFind}~\citep{Springel2001}. For a given host, the simulated halo population is randomly projected onto a 2D plane and systems within the projected virial radius are classified as satellites~\citep{Li2026a}. For direct comparison with the Merian sample, we apply the SHMR from \citet{Rodriguez2017} to estimate each host's $M_\star$ based on their halo mass and select all within the $9.5 < \log[M_{\star,\text{host}}/M_\odot] < 10.5$ range. We then estimate each subhalo's $M_\star$ with the SHMR from \citet{Nadler2020} based on their peak mass, sampling $100$ times from within its scatter, and select satellites with stellar mass $>10^{7.5} \ \rm M_{\odot}$. While the median values of the Merian data fall close or below the medians from TNG50, all bins have an $84^{\rm th}$ percentile above the theoretical $1\sigma$ boundary of TNG50. The most massive sub-MW-mass bin in particular shows an $84^{\rm th}$ percentile of $N_{\rm sat}>2$, while the corresponding host mass in TNG50 predicts $1$ satellite. This could indicate an overabundance of rich groups corresponding to a ``too many satellites'' problem at this mass range. However, these results may be influenced by high background contamination in the Merian sample and uncertainties in the SHMR. Mainly, any environmental dependence of the SHMR would shift the results predicted with the \citet{Nadler2020} relation, which is derived from MW satellites. Furthermore, Figure~\ref{fig:selection} illustrates how the host sample is incomplete at lower masses, which biases the results of each bin to the higher-mass end. This effect is not taken into account in the TNG50 data.

\subsection{Satellite radial distribution}

This section explores the radial distribution of satellites around the sub-MW-mass hosts. The spatial distribution of satellites around hosts is determined by the underlying dark matter distribution, but also processes such as dynamical friction and tidal stripping~\citep{Samuel2020,Webb2020,Green2021}. This thus represents another observable that can be directly compared to theoretical predictions of structure formation, possibly unveiling contradictions with the $\Lambda$CDM model~\citep{Nadler2021,Lovell2021}.

Figure~\ref{fig:Rhost} shows the cumulative radial distribution of satellites. For a given host, at each radial bin, we subtract the estimated background based on the area contained. We then take the average over the radial profiles for all $185$ hosts with total $N_{\rm sat}>0$. The $x$-axis is normalized by the virial radius of each host and the $y$-axis is normalized by the background-subtracted number of satellites within $R_{\text{vir}}$. We then bootstrap the distribution, resampling $185$ hosts at each iteration and computing the radial profile for the sample, then take the mean profile and $68\%$ containment range over $100$ iterations.

On the left panel, we compare our Merian sample with the MW, the ELVES satellite sample, and the MW-mass host Merian sample from \citet{Pan2026}. For the MW, we use the radial distributions for the satellite galaxies given by \citet{McConnachie.2012} and generate projected 2D distributions based on mock observations along $5000$ random lines of sight. The dashed-dotted line corresponds to the median from these projections and the shaded bands to the $1\sigma$ range. We include all satellites, including in the ultrafaint range, as a cut such as $M_{\star,\rm sat}>10^{7.5} \ \text{M}_\odot$ would limit statistics significantly, with only $3$ satellites. For the ELVES sample, we only include systems with coverage radius $>0.9\,R_{\rm vir}$ to ensure completeness and limit satellite masses to $M_{\star,\rm sat}>10^{7.5} \ \rm M_\odot$ for a closer comparison to our sample. The MW-mass Merian sample assumes the local background estimate described in Section~\ref{subsec:abundance} and is completeness-corrected assuming the quenched fraction~($f_{\rm quenched}$) from SAGA IV~\citep{Geha.etal.2024}. The ELVES sample includes many hosts ranging from the defined sub-MW-mass range to above MW-mass. The cumulative radial distributions for the ELVES and sub-MW-mass Merian samples largely agree, with slight deviations at low beings likely being attributed to the incompleteness of the Merian sample at low $d_{\rm proj}$. Despite spanning an entirely different mass range, the MW-mass sample approaches the sub-MW-mass profile when the $x$-axis is normalized by $R_{\rm vir}$. These results hint at a universal mean radial profile largely independent of host mass, which agrees with observations in the dwarf regime~\citep{Li2026b}. Meanwhile, the MW satellites are significantly more centrally concentrated than all other distributions, including the Merian MW-mass hosts. This is likely due to the influence from the Large Magellanic Cloud~\citep{DSouza2021,Samuel2021,Pan2026,Tan2026}. The inclusion of ultrafaint dwarfs in the MW also biases the distribution.

The right panel shows the theoretical predictions for the cumulative radial distribution for satellites with $M_{\rm \star, sat}>10^{7.5} \ \rm M_\odot$ from the TNG50 cosmological simulation for the sub-MW-mass range. The same bootstrapping is performed with the TNG50 sample, resampling $185$ hosts at each of $100$ iterations, to estimate the mean and $68\%$ percentile range, shown as a purple line and band, respectively. The Merian sample largely agrees with the TNG50 predictions, with both bands overlapping for most bins. The two samples deviate slightly at the lowest bins, which is likely due to the incompleteness of the Merian sample. Since the Merian survey targets primarily star-forming satellites, we miss any quenched systems, which are more common within the inner regions of the host. While the $f_{\rm quenched}$ of satellites around sub-MW-mass hosts is unknown, we show the shift in the distribution assuming the $f_{\rm quenched}$ for MW-mass hosts from SAGA IV~\citep{Geha.etal.2024}. We interpolate the $f_{\rm quenched}-d_{\rm proj}$ distribution to each bin center, and divide the background-subtracted radial distribution for each host by $(1-f_{\rm quenched})$. This introduces a central enhancement closer to the one predicted by TNG50. The Merian sample is also expected to be incomplete at $d_{\rm proj}<0.1R_{\rm vir}$, as it is difficult to identify faint sources near the bright host.

\section{DISCUSSION AND SUMMARY} \label{sec:conclusions}

Satellite dwarf galaxies represent an important test for models of hierarchical galaxy formation. The Merian survey presents a significantly larger statistical sample of bright, star-forming satellite dwarfs than previous surveys such as ELVES and SAGA. In particular, this study presents a large statistical census of sub-MW-mass hosts. 
Prior statistical work has concentrated on two ends of the host-mass scale: the dwarf regime~\citep{Hunter2025,Li2025,Li.etal.2025,Li2026b} and the MW-mass regime~\citep{McConnachie.2012, Savino2025, Geha2026}. With $148 \pm 7$ satellite candidates around 357 hosts, this work provides the first large sample in the intervening $10^{9.5}-10^{10.5} \ \rm M_\odot$ range, where hierarchical models predict satellite abundance to be a steep function of host mass.

Figure \ref{fig:Nsat_Nsys} shows the distribution of $N_{\text{sat}}$ for our sub-MW-mass systems, and compares the results to those of MW-like hosts~\citep{Pan2026}. We find a higher percentage of systems with low satellite counts, reflecting the expected decrease in satellite richness. In $\Lambda$CDM, the subhalo mass function is approximately self-similar, so the number of satellites above a fixed stellar-mass threshold should fall with host mass roughly in proportion to the host-to-satellite mass ratio; the trend we see is qualitatively consistent with this expectation. Figures~\ref{fig:Mstar} and \ref{fig:violins} also show that, at our mass range, there is a correlation between median satellite abundance and $M_{\star,\text{host}}$, $(g-i)_{\rm host}$ and $M_{\text{tot}}$, with a stronger correlation of the $84^{\rm th}$ percentiles. A similar increase with host mass has been found for H~\textsc{i}-bearing satellites selected around dwarf- to MW-mass hosts by \citet{Zhu2025}.

Figure \ref{fig:Rhost} shows the cumulative radial distribution for the satellites in our sample compared to other systems. When $d_{\rm proj}$ is normalized by $R_{\rm vir}$, the distributions for the sub-MW- and MW-mass hosts overlap. Satellites are expected to trace the host halo's mass distribution, so this result is consistent with $\Lambda$CDM predictions that halos across these mass ranges have similar concentrations~\citep{Bullok2001,Diemer2019}. The Merian data largely agree with the confirmed satellites around ELVES systems. The radial distribution also agrees with theoretical predictions in TNG50 for most radial bins. Results from the TNG50 simulation may be influenced by non-physical effects such as numerical stripping~\citep{Bosch2018}. While these effects are expected to be reduced at our mass range, they may still play a role in shaping the satellites in the inner regions of the host halo. Our observed sample is also incomplete within this low $d_{\rm proj}/R_{\rm vir}$ region, so both distributions may have similar shifts, causing this apparent agreement. The survey is primarily sensitive to star-forming satellites and thus may not be representative of the complete sample. Small deviations in radial profiles may thus provide relevant information on the quenching of satellites around sub-MW-mass hosts should the total satellite population be independently constrained by other surveys.

The large sample size and precise photometric redshift measurements of the Merian survey allow it to provide great insights to the study of satellites. Limitations on the completeness of lower-mass satellites and the scarcity of sub-MW-mass hosts in public group catalogs make it difficult to push the sample to lower host masses. However, our sample of 357 sub-MW-mass hosts still represents a great leap compared to previous surveys. Studies such as ELVES-Dwarf continue to provide robust measurements of satellite systems, with rigorous membership classification. The large statistical approach of our survey allows for a complementary population-level analysis across host masses with less rigorous constraints for individual systems. Another statistical survey, the DESI Bright Galaxy Survey~(BGS; \citealt{Wang.etal.2025}), recently measured the luminosity and stellar mass functions of photometric satellites around a broad range of isolated central galaxy masses~($7.1<\log_{10}(M_\star/\text{M}_\odot)<11.7$, extending to low magnitudes of $M_{r,\rm sat}\mathord{\sim}-7$. DESI BGS found that the faint-end slope of the satellite stellar mass function becomes steeper around lower-mass hosts. The Merian sample is complementary to this survey, as it focuses on the bright-end of the satellite population, which provides a robust picture of individual satellite properties and their radial distribution. The Merian survey will thus contribute to constraints in models of galaxy formation and evolution.

The present analysis demonstrates the power of Merian for constructing large statistical samples of satellite galaxies around a wide range of hosts, extending satellite studies into a host-mass regime that has previously been difficult to probe systematically. An important next step is spectroscopic follow-up of the satellite candidates. Targeted follow-up of the apparently richest systems would be particularly valuable for determining whether these systems represent genuinely satellite-rich halos. Looking further ahead, the Nancy Grace Roman Space Telescope~\citep{Spergel2013} and Euclid~\citep{Euclid2025} will combine wide-area imaging with the spatial resolution needed to obtain direct distance constraints for satellite candidates through, for example, surface brightness fluctuations, around much larger samples of nearby hosts. These measurements will complement the statistical approach demonstrated here by enabling direct membership determination and characterization of satellite populations on a system-by-system basis.
 
\begin{acknowledgments}

The authors would like to acknowledge helpful conversations and feedback from Pedro Cataldi, Akaxia Cruz, Cristiano Galbiati, Mariangela Lisanti, and Aviram Uri. They also acknowledge Tomer Romashenko for help with the visual inspection of the satellite candidates. 

The authors acknowledge the work reported in this paper was substantially performed using the Princeton Research Computing resources at Princeton University, a consortium of groups led by the Princeton Institute for Computational Science and Engineering (PICSciE) and the Office of Information Technology’s Research Computing.

The authors acknowledge support from Grant No. 2506292 from the United States National Science Foundation (NSF).

SD acknowledges funding and support from the European Research Council (ERC) under the European Union’s Horizon 2020 and Horizon Europe research and innovation programme (grant agreement number 101165294: HiddenUniverse), the Azrieli Foundation, and the Israel Science Foundation (ISF) under grant number 0603224992.

AL is supported by the National Science Foundation under Grant No. 2510898.

AHGP acknowledges support from U.S. National Science Foundation grant AST-2510899.

SH acknowledges support from the National Natural Science Foundation of China Grant No. 12273015 \& No. 12433003. 

\end{acknowledgments}

\software{\textsc{Jupyter~\citep{Kluyver2016jupyter}, \textsc{matplotlib}~\citep{Hunter:2007}, \textsc{NumPy}~\citep{harris2020array}, astropy \citep{2013A&A...558A..33A,2018AJ....156..123A}, \texttt{Halotools} \citep{Hearin.etal.2017}
          }}



\bibliography{references}{}

@ARTICLE{Tan2026,
       author = {{Tan}, C.~Y. and {Drlica-Wagner}, A. and {Pace}, A.~B. and {Cerny}, W. and {Nadler}, E.~O. and {Doliva-Dolinsky}, A. and {Anbajagane}, D. and {Li}, T.~S. and {Simon}, J.~D. and {Vivas}, A.~K. and {Walker}, A.~R. and {Adam{\'o}w}, M. and {Bechtol}, K. and {Carlin}, J.~L. and {Casey}, Q.~O. and {Chang}, C. and {Chaturvedi}, A. and {Cheng}, T.-Y. and {Chiti}, A. and {Choi}, Y. and {Crnojevi{\'c}}, D. and {Ferguson}, P.~S. and {Gruendl}, R.~A. and {Ji}, A.~P. and {Limberg}, G. and {Medina}, G.~E. and {Mutlu-Pakdil}, B. and {No{\"e}l}, N.~E.~D. and {Overdeck}, K. and {Placco}, V.~M. and {Riley}, A.~H. and {Sand}, D.~J. and {Sharp}, J. and {Sherman}, N.~F. and {Stringfellow}, G.~S. and {Wechsler}, R.~H. and {Aguena}, M. and {Allam}, S. and {Alves}, O. and {Bacon}, D. and {Brooks}, D. and {Burke}, D.~L. and {Camilleri}, R. and {Carballo-Bello}, J.~A. and {Carnero Rosell}, A. and {Carretero}, J. and {da Costa}, L.~N. and {da Silva Pereira}, M.~E. and {Davis}, T.~M. and {de Vicente}, J. and {Desai}, S. and {Everett}, S. and {Flaugher}, B. and {Frieman}, J. and {Garc{\'\i}a-Bellido}, J. and {Gruen}, D. and {Gutierrez}, G. and {Herner}, K. and {Hinton}, S.~R. and {Hollowood}, D.~L. and {James}, D.~J. and {Kuehn}, K. and {Lahav}, O. and {Lee}, S. and {Marshall}, J.~L. and {Mart{\'\i}nez-V{\'a}zquez}, C.~E. and {Massana}, P. and {Mena-Fern{\'a}ndez}, J. and {Miquel}, R. and {Muir}, J. and {Myles}, J. and {Ogando}, R.~L.~C. and {Plazas Malag{\'o}n}, A.~A. and {Porredon}, A. and {Sanchez}, E. and {Sanchez Cid}, D. and {Sevilla-Noarbe}, I. and {Smith}, M. and {Suchyta}, E. and {Swanson}, M.~E.~C. and {To}, C. and {Tollerud}, E.~J. and {Tucker}, D.~L. and {Vikram}, V. and {Weaverdyck}, N. and {Yamamoto}, M. and {Zenteno}, A. and {Delve Collaboration} and {Des Collaboration}},
        title = "{DELVE Milky Way Satellite Galaxy Census. I. Satellite Population and Survey Selection Function in DES, DELVE, and Pan-STARRS}",
      journal = {\apj},
         year = 2026,
        month = mar,
       volume = {1000},
       number = {1},
          eid = {87},
        pages = {87},
          doi = {10.3847/1538-4357/ae4479},
archivePrefix = {arXiv},
       eprint = {2509.12313},
 primaryClass = {astro-ph.GA},
       adsurl = {https://ui.adsabs.harvard.edu/abs/2026ApJ..1000...87T}
}

@Article{Hunter:2007,
  Author    = {Hunter, J. D.},
  Title     = {Matplotlib: A 2D graphics environment},
  Journal   = {Computing in Science \& Engineering},
  Volume    = {9},
  Number    = {3},
  Pages     = {90--95},
  publisher = {IEEE COMPUTER SOC},
  doi       = {10.1109/MCSE.2007.55},
  year      = 2007
}

@Article{harris2020array,
 title         = {Array programming with {NumPy}},
 author        = {Charles R. Harris and K. Jarrod Millman and St{\'{e}}fan J.
                 van der Walt and Ralf Gommers and Pauli Virtanen and David
                 Cournapeau and Eric Wieser and Julian Taylor and Sebastian
                 Berg and Nathaniel J. Smith and Robert Kern and Matti Picus
                 and Stephan Hoyer and Marten H. van Kerkwijk and Matthew
                 Brett and Allan Haldane and Jaime Fern{\'{a}}ndez del
                 R{\'{i}}o and Mark Wiebe and Pearu Peterson and Pierre
                 G{\'{e}}rard-Marchant and Kevin Sheppard and Tyler Reddy and
                 Warren Weckesser and Hameer Abbasi and Christoph Gohlke and
                 Travis E. Oliphant},
 year          = {2020},
 month         = sep,
 journal       = {Nature},
 volume        = {585},
 number        = {7825},
 pages         = {357--362},
 doi           = {10.1038/s41586-020-2649-2},
 publisher     = {Springer Science and Business Media {LLC}},
 url           = {https://doi.org/10.1038/s41586-020-2649-2}
}

@article{Kluyver2016jupyter,
   author = {Thomas Kluyver and Benjamin Ragan-Kelley and Fernando Pérez and Brian Granger and Matthias Bussonnier and Jonathan Frederic and Kyle Kelley and Jessica Hamrick and Jason Grout and Sylvain Corlay and Paul Ivanov and Damián Avila and Safia Abdalla and Carol Willing},
   doi = {10.3233/978-1-61499-649-1-87},
   isbn = {9781614996484},
   journal = {Positioning and Power in Academic Publishing: Players, Agents and Agendas - Proceedings of the 20th International Conference on Electronic Publishing, ELPUB 2016},
   pages = {87-90},
   publisher = {IOS Press},
   title = {Jupyter Notebooks – a publishing format for reproducible computational workflows},
   url = {https://ebooks.iospress.nl/doi/10.3233/978-1-61499-649-1-87},
   year = {2016},
}

@ARTICLE{delosReyes2025,
       author = {{de los Reyes}, Mithi A.~C. and {Asali}, Yasmeen and {Wechsler}, Risa H. and {Geha}, Marla and {Mao}, Yao-Yuan and {Kado-Fong}, Erin and {Pucha}, Ragadeepika and {Grant}, William and {Gandhi}, Pratik J. and {Manwadkar}, Viraj and {Engelhardt}, Anna and {Munshi}, Ferah and {Wang}, Yunchong},
        title = "{Stellar Mass Calibrations for Local Low-mass Galaxies}",
      journal = {\apj},
         year = 2025,
        month = aug,
       volume = {989},
       number = {1},
          eid = {91},
        pages = {91},
          doi = {10.3847/1538-4357/ade4c5},
archivePrefix = {arXiv},
       eprint = {2409.03959},
 primaryClass = {astro-ph.GA},
       adsurl = {https://ui.adsabs.harvard.edu/abs/2025ApJ...989...91D}
}

@ARTICLE{Roediger2015,
       author = {{Roediger}, Joel C. and {Courteau}, St{\'e}phane},
        title = "{On the uncertainties of stellar mass estimates via colour measurements}",
      journal = {\mnras},
         year = 2015,
        month = sep,
       volume = {452},
       number = {3},
        pages = {3209-3225},
          doi = {10.1093/mnras/stv1499},
archivePrefix = {arXiv},
       eprint = {1507.03016},
 primaryClass = {astro-ph.GA},
       adsurl = {https://ui.adsabs.harvard.edu/abs/2015MNRAS.452.3209R}
}

@ARTICLE{Conroy2009,
       author = {{Conroy}, Charlie and {Gunn}, James E. and {White}, Martin},
        title = "{The Propagation of Uncertainties in Stellar Population Synthesis Modeling. I. The Relevance of Uncertain Aspects of Stellar Evolution and the Initial Mass Function to the Derived Physical Properties of Galaxies}",
      journal = {\apj},
         year = 2009,
        month = jul,
       volume = {699},
       number = {1},
        pages = {486-506},
          doi = {10.1088/0004-637X/699/1/486},
archivePrefix = {arXiv},
       eprint = {0809.4261},
 primaryClass = {astro-ph},
       adsurl = {https://ui.adsabs.harvard.edu/abs/2009ApJ...699..486C}
}

@ARTICLE{Cataldi2026,
       author = {{Cataldi}, P. and {Cristiani}, V. and {Rodriguez}, F. and {Taverna}, A. and {Artale}, M.~C. and {Levine}, B. and {Wu}, H. and {the LSST Dark Energy Science Collaboration}},
        title = "{Halo occupation distribution estimation performance for LSST data}",
      journal = {\aap},
         year = 2026,
        month = aug,
       volume = {712},
          eid = {A114},
        pages = {A114},
          doi = {10.1051/0004-6361/202659674},
archivePrefix = {arXiv},
       eprint = {2603.01978},
 primaryClass = {astro-ph.CO},
       adsurl = {https://ui.adsabs.harvard.edu/abs/2026A&A...712A.114C}
}

@ARTICLE{Li2026b,
      title={ELVES-Dwarf. II. A Systematic Search for Satellite Systems of Dwarf Galaxies in the Local Volume}, 
      author={Jiaxuan Li and Jenny E. Greene and Shany Danieli and Scott G. Carlsten and John Moustakas and Marla Geha and Masayuki Tanaka and Fangzhou Jiang and Ping Chen and Sufia Birmingham},
      year={2026},
      eprint={2609.00283},
      archivePrefix={arXiv},
      primaryClass={astro-ph.GA},
      url={https://arxiv.org/abs/2609.00283}, 
}

@ARTICLE{Dekker2025,
       author = {{Dekker}, Ariane and {Kravtsov}, Andrey},
        title = "{Constraints on blue and red tilted primordial power spectra using dwarf galaxy properties}",
      journal = {\prd},
         year = 2025,
        month = mar,
       volume = {111},
       number = {6},
          eid = {063516},
        pages = {063516},
          doi = {10.1103/PhysRevD.111.063516},
archivePrefix = {arXiv},
       eprint = {2407.04198},
 primaryClass = {astro-ph.CO},
       adsurl = {https://ui.adsabs.harvard.edu/abs/2025PhRvD.111f3516D}
}

@ARTICLE{Zolotov2012,
       author = {{Zolotov}, Adi and {Brooks}, Alyson M. and {Willman}, Beth and {Governato}, Fabio and {Pontzen}, Andrew and {Christensen}, Charlotte and {Dekel}, Avishai and {Quinn}, Tom and {Shen}, Sijing and {Wadsley}, James},
        title = "{Baryons Matter: Why Luminous Satellite Galaxies have Reduced Central Masses}",
      journal = {\apj},
         year = 2012,
        month = dec,
       volume = {761},
       number = {1},
          eid = {71},
        pages = {71},
          doi = {10.1088/0004-637X/761/1/71},
archivePrefix = {arXiv},
       eprint = {1207.0007},
 primaryClass = {astro-ph.CO},
       adsurl = {https://ui.adsabs.harvard.edu/abs/2012ApJ...761...71Z}
}

@ARTICLE{Brooks2014,
       author = {{Brooks}, Alyson M. and {Zolotov}, Adi},
        title = "{Why Baryons Matter: The Kinematics of Dwarf Spheroidal Satellites}",
      journal = {\apj},
         year = 2014,
        month = may,
       volume = {786},
       number = {2},
          eid = {87},
        pages = {87},
          doi = {10.1088/0004-637X/786/2/87},
archivePrefix = {arXiv},
       eprint = {1207.2468},
 primaryClass = {astro-ph.CO},
       adsurl = {https://ui.adsabs.harvard.edu/abs/2014ApJ...786...87B}
}

@ARTICLE{Brooks2013,
       author = {{Brooks}, Alyson M. and {Kuhlen}, Michael and {Zolotov}, Adi and {Hooper}, Dan},
        title = "{A Baryonic Solution to the Missing Satellites Problem}",
      journal = {\apj},
         year = 2013,
        month = mar,
       volume = {765},
       number = {1},
          eid = {22},
        pages = {22},
          doi = {10.1088/0004-637X/765/1/22},
archivePrefix = {arXiv},
       eprint = {1209.5394},
 primaryClass = {astro-ph.CO},
       adsurl = {https://ui.adsabs.harvard.edu/abs/2013ApJ...765...22B}
}

@ARTICLE{Esteban2024,
       author = {{Esteban}, Ivan and {Peter}, Annika H.~G. and {Kim}, Stacy Y.},
        title = "{Milky Way satellite velocities reveal the dark matter power spectrum at small scales}",
      journal = {\prd},
         year = 2024,
        month = dec,
       volume = {110},
       number = {12},
          eid = {123013},
        pages = {123013},
          doi = {10.1103/PhysRevD.110.123013},
archivePrefix = {arXiv},
       eprint = {2306.04674},
 primaryClass = {astro-ph.CO},
       adsurl = {https://ui.adsabs.harvard.edu/abs/2024PhRvD.110l3013E}
}

@ARTICLE{Kim2021,
       author = {{Kim}, Stacy Y. and {Peter}, Annika H.~G.},
        title = "{The Milky Way satellite velocity function is a sharp probe of small-scale structure problems}",
      journal = {arXiv e-prints},
         year = 2021,
        month = jun,
          eid = {arXiv:2106.09050},
        pages = {arXiv:2106.09050},
          doi = {10.48550/arXiv.2106.09050},
archivePrefix = {arXiv},
       eprint = {2106.09050},
 primaryClass = {astro-ph.GA},
       adsurl = {https://ui.adsabs.harvard.edu/abs/2021arXiv210609050K}
}

@ARTICLE{Paudel2026,
       author = {{Paudel}, Sanjaya and {Sabiu}, Cristiano G. and {Yoon}, Suk-Jin and {Nidhi Chhatkuli}, Daya and {Yoo}, Jaewon and {Pokhrel}, Nau Raj},
        title = "{Groups of Dwarf Galaxies in the Local Universe}",
      journal = {arXiv e-prints},
         year = 2026,
        month = jun,
          eid = {arXiv:2606.16299},
        pages = {arXiv:2606.16299},
          doi = {10.48550/arXiv.2606.16299},
archivePrefix = {arXiv},
       eprint = {2606.16299},
 primaryClass = {astro-ph.GA},
       adsurl = {https://ui.adsabs.harvard.edu/abs/2026arXiv260616299P}
}

@ARTICLE{Roberts2021,
       author = {{Roberts}, Daniella M. and {Nierenberg}, Anna M. and {Peter}, Annika H.~G.},
        title = "{The luminosity functions and redshift evolution of satellites of low-mass galaxies in the COSMOS survey}",
      journal = {\mnras},
         year = 2021,
        month = mar,
       volume = {502},
       number = {1},
        pages = {1205-1217},
          doi = {10.1093/mnras/stab069},
archivePrefix = {arXiv},
       eprint = {2008.05479},
 primaryClass = {astro-ph.GA},
       adsurl = {https://ui.adsabs.harvard.edu/abs/2021MNRAS.502.1205R}
}

@ARTICLE{Nierenberg2012,
       author = {{Nierenberg}, A.~M. and {Auger}, M.~W. and {Treu}, T. and {Marshall}, P.~J. and {Fassnacht}, C.~D. and {Busha}, Michael T.},
        title = "{Luminous Satellites. II. Spatial Distribution, Luminosity Function, and Cosmic Evolution}",
      journal = {\apj},
         year = 2012,
        month = jun,
       volume = {752},
       number = {2},
          eid = {99},
        pages = {99},
          doi = {10.1088/0004-637X/752/2/99},
archivePrefix = {arXiv},
       eprint = {1202.2125},
 primaryClass = {astro-ph.GA},
       adsurl = {https://ui.adsabs.harvard.edu/abs/2012ApJ...752...99N}
}

@ARTICLE{Geha2026,
       author = {{Geha}, Marla},
        title = "{The Keck/DEIMOS Stellar Archive: II. Dynamical Masses and Metallicities for a Uniform Sample of Milky Way Satellites}",
      journal = {arXiv e-prints},
         year = 2026,
        month = feb,
          eid = {arXiv:2602.10202},
        pages = {arXiv:2602.10202},
          doi = {10.48550/arXiv.2602.10202},
archivePrefix = {arXiv},
       eprint = {2602.10202},
 primaryClass = {astro-ph.GA},
       adsurl = {https://ui.adsabs.harvard.edu/abs/2026arXiv260210202G}
}

@Article{Zhu2025,
  author        = {{Zhu}, Jingyao and {Asali}, Yasmeen and {Putman}, Mary and
                  {Westmeier}, Tobias and {de Blok}, W.~J. G and {Catinella},
                  Barbara and {Deg}, Nathan and {For}, Bi-Qing and {Kleiner},
                  Dane and {Lee-Waddell}, Karen and {Maccagni}, Filippo and
                  {Pisano}, D.~J. and {Shen}, Austin X. and {Spekkens},
                  Kristine and {Staveley-Smith}, Lister},
  title         = "{Baryonic Masses and Properties of Gaseous Satellite
                  Galaxies}",
  journal       = {arXiv e-prints},
  year          = 2025,
  month         = oct,
  eid           = {arXiv:2510.27019},
  pages         = {arXiv:2510.27019},
  doi           = {10.48550/arXiv.2510.27019},
  archiveprefix = {arXiv},
  eprint        = {2510.27019},
  primaryclass  = {astro-ph.GA},
  adsurl        = {https://ui.adsabs.harvard.edu/abs/2025arXiv251027019Z}
}

@ARTICLE{Ilbert2009,
       author = {{Ilbert}, O. and {Capak}, P. and {Salvato}, M. and {Aussel}, H. and {McCracken}, H.~J. and {Sanders}, D.~B. and {Scoville}, N. and {Kartaltepe}, J. and {Arnouts}, S. and {Le Floc'h}, E. and {Mobasher}, B. and {Taniguchi}, Y. and {Lamareille}, F. and {Leauthaud}, A. and {Sasaki}, S. and {Thompson}, D. and {Zamojski}, M. and {Zamorani}, G. and {Bardelli}, S. and {Bolzonella}, M. and {Bongiorno}, A. and {Brusa}, M. and {Caputi}, K.~I. and {Carollo}, C.~M. and {Contini}, T. and {Cook}, R. and {Coppa}, G. and {Cucciati}, O. and {de la Torre}, S. and {de Ravel}, L. and {Franzetti}, P. and {Garilli}, B. and {Hasinger}, G. and {Iovino}, A. and {Kampczyk}, P. and {Kneib}, J.-P. and {Knobel}, C. and {Kovac}, K. and {Le Borgne}, J.~F. and {Le Brun}, V. and {Le F{\`e}vre}, O. and {Lilly}, S. and {Looper}, D. and {Maier}, C. and {Mainieri}, V. and {Mellier}, Y. and {Mignoli}, M. and {Murayama}, T. and {Pell{\`o}}, R. and {Peng}, Y. and {P{\'e}rez-Montero}, E. and {Renzini}, A. and {Ricciardelli}, E. and {Schiminovich}, D. and {Scodeggio}, M. and {Shioya}, Y. and {Silverman}, J. and {Surace}, J. and {Tanaka}, M. and {Tasca}, L. and {Tresse}, L. and {Vergani}, D. and {Zucca}, E.},
        title = "{Cosmos Photometric Redshifts with 30-Bands for 2-deg$^{2}$}",
      journal = {\apj},
         year = 2009,
        month = jan,
       volume = {690},
       number = {2},
        pages = {1236-1249},
          doi = {10.1088/0004-637X/690/2/1236},
archivePrefix = {arXiv},
       eprint = {0809.2101},
 primaryClass = {astro-ph},
       adsurl = {https://ui.adsabs.harvard.edu/abs/2009ApJ...690.1236I}
}

@ARTICLE{Euclid2025,
       author = {{Euclid Collaboration} and {Mellier}, Y. and {Abdurro'uf} and {Acevedo Barroso}, J.~A. and {Ach{\'u}carro}, A. and {Adamek}, J. and {Adam}, R. and {Addison}, G.~E. and {Aghanim}, N. and {Aguena}, M. and {Ajani}, V. and {Akrami}, Y. and {Al-Bahlawan}, A. and {Alavi}, A. and {Albuquerque}, I.~S. and {Alestas}, G. and {Alguero}, G. and {Allaoui}, A. and {Allen}, S.~W. and {Allevato}, V. and {Alonso-Tetilla}, A.~V. and {Altieri}, B. and {Alvarez-Candal}, A. and {Alvi}, S. and {Amara}, A. and {Amendola}, L. and {Amiaux}, J. and {Andika}, I.~T. and {Andreon}, S. and {Andrews}, A. and {Angora}, G. and {Angulo}, R.~E. and {Annibali}, F. and {Anselmi}, A. and {Anselmi}, S. and {Arcari}, S. and {Archidiacono}, M. and {Aric{\`o}}, G. and {Arnaud}, M. and {Arnouts}, S. and {Asgari}, M. and {Asorey}, J. and {Atayde}, L. and {Atek}, H. and {Atrio-Barandela}, F. and {Aubert}, M. and {Aubourg}, E. and {Auphan}, T. and {Auricchio}, N. and {Aussel}, B. and {Aussel}, H. and {Avelino}, P.~P. and {Avgoustidis}, A. and {Avila}, S. and {Awan}, S. and {Azzollini}, R. and {Baccigalupi}, C. and {Bachelet}, E. and {Bacon}, D. and {Baes}, M. and {Bagley}, M.~B. and {Bahr-Kalus}, B. and {Balaguera-Antolinez}, A. and {Balbinot}, E. and {Balcells}, M. and {Baldi}, M. and {Baldry}, I. and {Balestra}, A. and {Ballardini}, M. and {Ballester}, O. and {Balogh}, M. and {Ba{\~n}ados}, E. and {Barbier}, R. and {Bardelli}, S. and {Baron}, M. and {Barreiro}, T. and {Barrena}, R. and {Barriere}, J.-C. and {Barros}, B.~J. and {Barthelemy}, A. and {Bartolo}, N. and {Basset}, A. and {Battaglia}, P. and {Battisti}, A.~J. and {Baugh}, C.~M. and {Baumont}, L. and {Bazzanini}, L. and {Beaulieu}, J.-P. and {Beckmann}, V. and {Belikov}, A.~N. and {Bel}, J. and {Bellagamba}, F. and {Bella}, M. and {Bellini}, E. and {Benabed}, K. and {Bender}, R. and {Benevento}, G. and {Bennett}, C.~L. and {Benson}, K. and {Bergamini}, P. and {Bermejo-Climent}, J.~R. and {Bernardeau}, F. and {Bertacca}, D. and {Berthe}, M. and {Berthier}, J. and {Bethermin}, M. and {Beutler}, F. and {Bevillon}, C. and {Bhargava}, S. and {Bhatawdekar}, R. and {Bianchi}, D. and {Bisigello}, L. and {Biviano}, A. and {Blake}, R.~P. and {Blanchard}, A. and {Blazek}, J. and {Blot}, L. and {Bosco}, A. and {Bodendorf}, C. and {Boenke}, T. and {B{\"o}hringer}, H. and {Boldrini}, P. and {Bolzonella}, M. and {Bonchi}, A. and {Bonici}, M. and {Bonino}, D. and {Bonino}, L. and {Bonvin}, C. and {Bon}, W. and {Booth}, J.~T. and {Borgani}, S. and {Borlaff}, A.~S. and {Borsato}, E. and {Bose}, B. and {Botticella}, M.~T. and {Boucaud}, A. and {Bouche}, F. and {Boucher}, J.~S. and {Boutigny}, D. and {Bouvard}, T. and {Bouwens}, R. and {Bouy}, H. and {Bowler}, R.~A.~A. and {Bozza}, V. and {Bozzo}, E. and {Branchini}, E. and {Brando}, G. and {Brau-Nogue}, S. and {Brekke}, P. and {Bremer}, M.~N. and {Brescia}, M. and {Breton}, M.-A. and {Brinchmann}, J. and {Brinckmann}, T. and {Brockley-Blatt}, C. and {Brodwin}, M. and {Brouard}, L. and {Brown}, M.~L. and {Bruton}, S. and {Bucko}, J. and {Buddelmeijer}, H. and {Buenadicha}, G. and {Buitrago}, F. and {Burger}, P. and {Burigana}, C. and {Busillo}, V. and {Busonero}, D. and {Cabanac}, R. and {Cabayol-Garcia}, L. and {Cagliari}, M.~S. and {Caillat}, A. and {Caillat}, L. and {Calabrese}, M. and {Calabro}, A. and {Calderone}, G. and {Calura}, F. and {Camacho Quevedo}, B. and {Camera}, S. and {Campos}, L. and {Ca{\~n}as-Herrera}, G. and {Candini}, G.~P. and {Cantiello}, M. and {Capobianco}, V. and {Cappellaro}, E. and {Cappelluti}, N. and {Cappi}, A. and {Caputi}, K.~I. and {Cara}, C. and {Carbone}, C. and {Cardone}, V.~F. and {Carella}, E. and {Carlberg}, R.~G. and {Carle}, M. and {Carminati}, L. and {Caro}, F. and {Carrasco}, J.~M. and {Carretero}, J. and {Carrilho}, P. and {Carron Duque}, J. and {Carry}, B.},
        title = "{Euclid: I. Overview of the Euclid mission}",
      journal = {\aap},
         year = 2025,
        month = may,
       volume = {697},
          eid = {A1},
        pages = {A1},
          doi = {10.1051/0004-6361/202450810},
archivePrefix = {arXiv},
       eprint = {2405.13491},
 primaryClass = {astro-ph.CO},
       adsurl = {https://ui.adsabs.harvard.edu/abs/2025A&A...697A...1E}
}

@ARTICLE{Spergel2013,
       author = {{Spergel}, D. and {Gehrels}, N. and {Breckinridge}, J. and {Donahue}, M. and {Dressler}, A. and {Gaudi}, B.~S. and {Greene}, T. and {Guyon}, O. and {Hirata}, C. and {Kalirai}, J. and {Kasdin}, N.~J. and {Moos}, W. and {Perlmutter}, S. and {Postman}, M. and {Rauscher}, B. and {Rhodes}, J. and {Wang}, Y. and {Weinberg}, D. and {Centrella}, J. and {Traub}, W. and {Baltay}, C. and {Colbert}, J. and {Bennett}, D. and {Kiessling}, A. and {Macintosh}, B. and {Merten}, J. and {Mortonson}, M. and {Penny}, M. and {Rozo}, E. and {Savransky}, D. and {Stapelfeldt}, K. and {Zu}, Y. and {Baker}, C. and {Cheng}, E. and {Content}, D. and {Dooley}, J. and {Foote}, M. and {Goullioud}, R. and {Grady}, K. and {Jackson}, C. and {Kruk}, J. and {Levine}, M. and {Melton}, M. and {Peddie}, C. and {Ruffa}, J. and {Shaklan}, S.},
        title = "{Wide-Field InfraRed Survey Telescope-Astrophysics Focused Telescope Assets WFIRST-AFTA Final Report}",
      journal = {arXiv e-prints},
         year = 2013,
        month = may,
          eid = {arXiv:1305.5422},
        pages = {arXiv:1305.5422},
          doi = {10.48550/arXiv.1305.5422},
archivePrefix = {arXiv},
       eprint = {1305.5422},
 primaryClass = {astro-ph.IM},
       adsurl = {https://ui.adsabs.harvard.edu/abs/2013arXiv1305.5422S}
}

@ARTICLE{Danieli2023,
       author = {{Danieli}, Shany and {Greene}, Jenny E. and {Carlsten}, Scott and {Jiang}, Fangzhou and {Beaton}, Rachael and {Goulding}, Andy D.},
        title = "{ELVES. IV. The Satellite Stellar-to-halo Mass Relation Beyond the Milky Way}",
      journal = {\apj},
         year = 2023,
        month = oct,
       volume = {956},
       number = {1},
          eid = {6},
        pages = {6},
          doi = {10.3847/1538-4357/acefbd},
archivePrefix = {arXiv},
       eprint = {2210.14233},
 primaryClass = {astro-ph.GA},
       adsurl = {https://ui.adsabs.harvard.edu/abs/2023ApJ...956....6D}
}

@ARTICLE{Lovell2021,
       author = {{Lovell}, Mark R. and {Cautun}, Marius and {Frenk}, Carlos S. and {Hellwing}, Wojciech A. and {Newton}, Oliver},
        title = "{The spatial distribution of Milky Way satellites, gaps in streams, and the nature of dark matter}",
      journal = {\mnras},
         year = 2021,
        month = nov,
       volume = {507},
       number = {4},
        pages = {4826-4839},
          doi = {10.1093/mnras/stab2452},
archivePrefix = {arXiv},
       eprint = {2104.03322},
 primaryClass = {astro-ph.GA},
       adsurl = {https://ui.adsabs.harvard.edu/abs/2021MNRAS.507.4826L}
}

@ARTICLE{Nadler2021,
       author = {{Nadler}, Ethan O. and {Banerjee}, Arka and {Adhikari}, Susmita and {Mao}, Yao-Yuan and {Wechsler}, Risa H.},
        title = "{The Effects of Dark Matter and Baryonic Physics on the Milky Way Subhalo Population in the Presence of the Large Magellanic Cloud}",
      journal = {\apjl},
         year = 2021,
        month = oct,
       volume = {920},
       number = {1},
          eid = {L11},
        pages = {L11},
          doi = {10.3847/2041-8213/ac29c1},
archivePrefix = {arXiv},
       eprint = {2109.12120},
 primaryClass = {astro-ph.CO},
       adsurl = {https://ui.adsabs.harvard.edu/abs/2021ApJ...920L..11N}
}

@ARTICLE{Green2021,
       author = {{Green}, Sheridan B. and {van den Bosch}, Frank C. and {Jiang}, Fangzhou},
        title = "{The tidal evolution of dark matter substructure - II. The impact of artificial disruption on subhalo mass functions and radial profiles}",
      journal = {\mnras},
         year = 2021,
        month = may,
       volume = {503},
       number = {3},
        pages = {4075-4091},
          doi = {10.1093/mnras/stab696},
archivePrefix = {arXiv},
       eprint = {2103.01227},
 primaryClass = {astro-ph.GA},
       adsurl = {https://ui.adsabs.harvard.edu/abs/2021MNRAS.503.4075G}
}

@ARTICLE{Webb2020,
       author = {{Webb}, Jeremy J. and {Bovy}, Jo},
        title = "{High-resolution simulations of dark matter subhalo disruption in a Milky-Way-like tidal field}",
      journal = {\mnras},
         year = 2020,
        month = nov,
       volume = {499},
       number = {1},
        pages = {116-128},
          doi = {10.1093/mnras/staa2852},
archivePrefix = {arXiv},
       eprint = {2006.06695},
 primaryClass = {astro-ph.GA},
       adsurl = {https://ui.adsabs.harvard.edu/abs/2020MNRAS.499..116W}
}

@ARTICLE{Samuel2020,
       author = {{Samuel}, Jenna and {Wetzel}, Andrew and {Tollerud}, Erik and {Garrison-Kimmel}, Shea and {Loebman}, Sarah and {El-Badry}, Kareem and {Hopkins}, Philip F. and {Boylan-Kolchin}, Michael and {Faucher-Gigu{\`e}re}, Claude-Andr{\'e} and {Bullock}, James S. and {Benincasa}, Samantha and {Bailin}, Jeremy},
        title = "{A profile in FIRE: resolving the radial distributions of satellite galaxies in the Local Group with simulations}",
      journal = {\mnras},
         year = 2020,
        month = jan,
       volume = {491},
       number = {1},
        pages = {1471-1490},
          doi = {10.1093/mnras/stz3054},
archivePrefix = {arXiv},
       eprint = {1904.11508},
 primaryClass = {astro-ph.GA},
       adsurl = {https://ui.adsabs.harvard.edu/abs/2020MNRAS.491.1471S}
}

@ARTICLE{Putman2012,
       author = {{Putman}, M.~E. and {Peek}, J.~E.~G. and {Joung}, M.~R.},
        title = "{Gaseous Galaxy Halos}",
      journal = {\araa},
         year = 2012,
        month = sep,
       volume = {50},
        pages = {491-529},
          doi = {10.1146/annurev-astro-081811-125612},
archivePrefix = {arXiv},
       eprint = {1207.4837},
 primaryClass = {astro-ph.GA},
       adsurl = {https://ui.adsabs.harvard.edu/abs/2012ARA&A..50..491P}
}

@ARTICLE{Weisz2013,
       author = {{Weisz}, Daniel R. and {Dolphin}, Andrew E. and {Skillman}, Evan D. and {Holtzman}, Jon and {Dalcanton}, Julianne J. and {Cole}, Andrew A. and {Neary}, Kyle},
        title = "{Comparing the ancient star formation histories of the Magellanic Clouds}",
      journal = {\mnras},
         year = 2013,
        month = may,
       volume = {431},
       number = {1},
        pages = {364-371},
          doi = {10.1093/mnras/stt165},
archivePrefix = {arXiv},
       eprint = {1301.7422},
 primaryClass = {astro-ph.CO},
       adsurl = {https://ui.adsabs.harvard.edu/abs/2013MNRAS.431..364W}
}

@ARTICLE{Sales2022,
       author = {{Sales}, Laura V. and {Wetzel}, Andrew and {Fattahi}, Azadeh},
        title = "{Baryonic solutions and challenges for cosmological models of dwarf galaxies}",
      journal = {Nature Astronomy},
         year = 2022,
        month = jun,
       volume = {6},
        pages = {897-910},
          doi = {10.1038/s41550-022-01689-w},
archivePrefix = {arXiv},
       eprint = {2206.05295},
 primaryClass = {astro-ph.GA},
       adsurl = {https://ui.adsabs.harvard.edu/abs/2022NatAs...6..897S}
}

@ARTICLE{Hunter2025,
       author = {{Hunter}, Laura Congreve and {Mutlu-Pakd{\.I}l}, Bur{\c{c}}{\.I}n and {Sand}, David J. and {Bennet}, Paul and {Khim}, Donghyeon J. and {Crnojevi{\'c}}, Denija and {Doliva-Dolinsky}, Amandine and {Durodola}, Emmanuel and {Fielder}, Catherine and {Goebel-Bain}, Rowan and {Jones}, Michael G. and {Karunakaran}, Ananthan and {Spekkens}, Kristine and {Zaritsky}, Dennis},
        title = "{Identifying Dwarfs of MC Analog GalaxiEs (ID-MAGE): The Search for Satellites around Low-mass Hosts}",
      journal = {\apj},
         year = 2025,
        month = aug,
       volume = {989},
       number = {1},
          eid = {58},
        pages = {58},
          doi = {10.3847/1538-4357/ade9a4},
archivePrefix = {arXiv},
       eprint = {2504.00253},
 primaryClass = {astro-ph.GA},
       adsurl = {https://ui.adsabs.harvard.edu/abs/2025ApJ...989...58H}
}

@ARTICLE{Doliva2025,
       author = {{Doliva-Dolinsky}, Amandine and {Mutlu-Pakdil}, Bur{\c{c}}in and {Crnojevi{\'c}}, Denija and {Anbajagane}, Dhayaa and {Carlin}, Jeffrey L. and {Medoff}, Jonah and {Sand}, David J. and {Tollerud}, Erik and {Lim}, Sungsoon and {Bennet}, Paul and {Drlica-Wagner}, Alex and {Fielder}, Catherine E. and {Hargis}, Jonathan R. and {Herron}, Kai and {Hunter}, Laura Congreve and {Jones}, Michael G. and {Karunakaran}, Ananthan and {Peter}, Annika H.~G. and {Romanowsky}, Aaron J. and {Spekkens}, Kristine and {Strader}, Jay and {Willman}, Beth and {Carballo-Bello}, Julio A. and {Cerny}, William and {Chaturvedi}, Astha and {Kallivayalil}, Nitya and {Mart{\'\i}nez-V{\'a}zquez}, Clara E. and {Medina}, Gustavo E. and {No{\"e}l}, Noelia E.~D. and {Pace}, Andrew B. and {Riley}, Alex H. and {Sakowska}, Joanna D. and {Smercina}, Adam and {Vivas}, Kathy and {Adam{\'o}w}, Monika and {Bom}, Clecio R. and {Choi}, Yumi and {Ferguson}, Peter S. and {Navabi}, Mahdieh and {Zenteno}, Alfredo and {Delve Collaboration}},
        title = "{The NGC 3109 Satellite System: The First Systematic Resolved Search for Dwarf Galaxies Around an SMC-mass Host}",
      journal = {\apj},
         year = 2025,
        month = aug,
       volume = {989},
       number = {1},
          eid = {21},
        pages = {21},
          doi = {10.3847/1538-4357/ade9b8},
archivePrefix = {arXiv},
       eprint = {2505.05570},
 primaryClass = {astro-ph.GA},
       adsurl = {https://ui.adsabs.harvard.edu/abs/2025ApJ...989...21D}
}

@ARTICLE{McNanna2024,
       author = {{McNanna}, M. and {Bechtol}, K. and {Mau}, S. and {Nadler}, E.~O. and {Medoff}, J. and {Drlica-Wagner}, A. and {Cerny}, W. and {Crnojevi{\'c}}, D. and {Mutlu-Pakd{\i}l}, B. and {Vivas}, A.~K. and {Pace}, A.~B. and {Carlin}, J.~L. and {Collins}, M.~L.~M. and {Ferguson}, P.~S. and {Mart{\'\i}nez-Delgado}, D. and {Mart{\'\i}nez-V{\'a}zquez}, C.~E. and {Noel}, N.~E.~D. and {Riley}, A.~H. and {Sand}, D.~J. and {Smercina}, A. and {Tollerud}, E. and {Wechsler}, R.~H. and {Abbott}, T.~M.~C. and {Aguena}, M. and {Alves}, O. and {Bacon}, D. and {Bom}, C.~R. and {Brooks}, D. and {Burke}, D.~L. and {Carballo-Bello}, J.~A. and {Carnero Rosell}, A. and {Carretero}, J. and {da Costa}, L.~N. and {Davis}, T.~M. and {de Vicente}, J. and {Diehl}, H.~T. and {Doel}, P. and {Ferrero}, I. and {Frieman}, J. and {Giannini}, G. and {Gruen}, D. and {Gutierrez}, G. and {Gruendl}, R.~A. and {Hinton}, S.~R. and {Hollowood}, D.~L. and {Honscheid}, K. and {James}, D.~J. and {Kuehn}, K. and {Marshall}, J.~L. and {Mena-Fern{\'a}ndez}, J. and {Miquel}, R. and {Pereira}, M.~E.~S. and {Pieres}, A. and {Malag{\'o}n}, A.~A. Plazas and {Sakowska}, J.~D. and {Sanchez}, E. and {Sanchez Cid}, D. and {Santiago}, B. and {Sevilla-Noarbe}, I. and {Smith}, M. and {Stringfellow}, G.~S. and {Suchyta}, E. and {Swanson}, M.~E.~C. and {Tarle}, G. and {Weaverdyck}, N. and {Wiseman}, P. and {DES Collaboration} and {DELVE Collaboration}},
        title = "{A Search for Faint Resolved Galaxies Beyond the Milky Way in DES Year 6: A New Faint, Diffuse Dwarf Satellite of NGC 55}",
      journal = {\apj},
         year = 2024,
        month = jan,
       volume = {961},
       number = {1},
          eid = {126},
        pages = {126},
          doi = {10.3847/1538-4357/ad07d0},
archivePrefix = {arXiv},
       eprint = {2309.04467},
 primaryClass = {astro-ph.GA},
       adsurl = {https://ui.adsabs.harvard.edu/abs/2024ApJ...961..126M}
}

@ARTICLE{Drlica2021,
       author = {{Drlica-Wagner}, A. and {Carlin}, J.~L. and {Nidever}, D.~L. and {Ferguson}, P.~S. and {Kuropatkin}, N. and {Adam{\'o}w}, M. and {Cerny}, W. and {Choi}, Y. and {Esteves}, J.~H. and {Mart{\'\i}nez-V{\'a}zquez}, C.~E. and {Mau}, S. and {Miller}, A.~E. and {Mutlu-Pakdil}, B. and {Neilsen}, E.~H. and {Olsen}, K.~A.~G. and {Pace}, A.~B. and {Riley}, A.~H. and {Sakowska}, J.~D. and {Sand}, D.~J. and {Santana-Silva}, L. and {Tollerud}, E.~J. and {Tucker}, D.~L. and {Vivas}, A.~K. and {Zaborowski}, E. and {Zenteno}, A. and {Abbott}, T.~M.~C. and {Allam}, S. and {Bechtol}, K. and {Bell}, C.~P.~M. and {Bell}, E.~F. and {Bilaji}, P. and {Bom}, C.~R. and {Carballo-Bello}, J.~A. and {Crnojevi{\'c}}, D. and {Cioni}, M.-R.~L. and {Diaz-Ocampo}, A. and {de Boer}, T.~J.~L. and {Erkal}, D. and {Gruendl}, R.~A. and {Hernandez-Lang}, D. and {Hughes}, A.~K. and {James}, D.~J. and {Johnson}, L.~C. and {Li}, T.~S. and {Mao}, Y.-Y. and {Mart{\'\i}nez-Delgado}, D. and {Massana}, P. and {McNanna}, M. and {Morgan}, R. and {Nadler}, E.~O. and {No{\"e}l}, N.~E.~D. and {Palmese}, A. and {Peter}, A.~H.~G. and {Rykoff}, E.~S. and {S{\'a}nchez}, J. and {Shipp}, N. and {Simon}, J.~D. and {Smercina}, A. and {Soares-Santos}, M. and {Stringfellow}, G.~S. and {Tavangar}, K. and {van der Marel}, R.~P. and {Walker}, A.~R. and {Wechsler}, R.~H. and {Wu}, J.~F. and {Yanny}, B. and {Fitzpatrick}, M. and {Huang}, L. and {Jacques}, A. and {Nikutta}, R. and {Scott}, A. and {Astro Data Lab}},
        title = "{The DECam Local Volume Exploration Survey: Overview and First Data Release}",
      journal = {\apjs},
         year = 2021,
        month = sep,
       volume = {256},
       number = {1},
          eid = {2},
        pages = {2},
          doi = {10.3847/1538-4365/ac079d},
archivePrefix = {arXiv},
       eprint = {2103.07476},
 primaryClass = {astro-ph.GA},
       adsurl = {https://ui.adsabs.harvard.edu/abs/2021ApJS..256....2D}
}

@ARTICLE{Davis2024,
       author = {{Davis}, A. Bianca and {Garling}, Christopher T. and {Nierenberg}, Anna M. and {Peter}, Annika H.~G. and {Sardone}, Amy and {Kochanek}, Christopher S. and {Leroy}, Adam K. and {Casey}, Kirsten J. and {Pogge}, Richard W. and {Roberts}, Daniella M. and {Sand}, David J. and {Greco}, Johnny P.},
        title = "{The LBT Satellites of Nearby Galaxies Survey (LBT-SONG): The Diffuse Satellite Population of Local Volume Hosts}",
      journal = {arXiv e-prints},
         year = 2024,
        month = sep,
          eid = {arXiv:2409.03999},
        pages = {arXiv:2409.03999},
          doi = {10.48550/arXiv.2409.03999},
archivePrefix = {arXiv},
       eprint = {2409.03999},
 primaryClass = {astro-ph.GA},
       adsurl = {https://ui.adsabs.harvard.edu/abs/2024arXiv240903999D}
}

@ARTICLE{Davis2021,
       author = {{Davis}, A. Bianca and {Nierenberg}, Anna M. and {Peter}, Annika H.~G. and {Garling}, Christopher T. and {Greco}, Johnny P. and {Kochanek}, Christopher S. and {Utomo}, Dyas and {Casey}, Kirsten J. and {Pogge}, Richard W. and {Roberts}, Daniella M. and {Sand}, David J. and {Sardone}, Amy},
        title = "{The LBT satellites of Nearby Galaxies Survey (LBT-SONG): the satellite population of NGC 628}",
      journal = {\mnras},
         year = 2021,
        month = jan,
       volume = {500},
       number = {3},
        pages = {3854-3869},
          doi = {10.1093/mnras/staa3246},
archivePrefix = {arXiv},
       eprint = {2003.08352},
 primaryClass = {astro-ph.GA},
       adsurl = {https://ui.adsabs.harvard.edu/abs/2021MNRAS.500.3854D}
}

@ARTICLE{Garling2021,
       author = {{Garling}, Christopher T. and {Peter}, Annika H.~G. and {Kochanek}, Christopher S. and {Sand}, David J. and {Crnojevi{\'c}}, Denija},
        title = "{A search for satellite galaxies of nearby star-forming galaxies with resolved stars in LBT-SONG}",
      journal = {\mnras},
         year = 2021,
        month = nov,
       volume = {507},
       number = {4},
        pages = {4764-4778},
          doi = {10.1093/mnras/stab2447},
archivePrefix = {arXiv},
       eprint = {2105.01082},
 primaryClass = {astro-ph.GA},
       adsurl = {https://ui.adsabs.harvard.edu/abs/2021MNRAS.507.4764G}
}

@INPROCEEDINGS{Carlin2025,
       author = {{Carlin}, Jeffrey and {Sand}, David and {Mutlu Pakdil}, Burcin and {Crnojevic}, Denija and {Peter}, Annika and {Doliva-Dolinsky}, Amandine and {Garling}, Christopher and {Brodie}, Jean and {Forbes}, Duncan and {Hargis}, Jonathan and {Romanowsky}, Aaron and {Spekkens}, Kristine and {Strader}, Jay and {Willman}, Beth},
        title = "{A census of dwarf galaxy satellites around LMC-mass galaxy NGC 2403}",
    booktitle = {American Astronomical Society Meeting Abstracts \#245},
         year = 2025,
       series = {American Astronomical Society Meeting Abstracts},
       volume = {245},
        month = jan,
          eid = {365.05},
        pages = {365.05},
       adsurl = {https://ui.adsabs.harvard.edu/abs/2025AAS...24536505C}
}

@ARTICLE{Carlin2016,
       author = {{Carlin}, Jeffrey L. and {Sand}, David J. and {Price}, Paul and {Willman}, Beth and {Karunakaran}, Ananthan and {Spekkens}, Kristine and {Bell}, Eric F. and {Brodie}, Jean P. and {Crnojevi{\'c}}, Denija and {Forbes}, Duncan A. and {Hargis}, Jonathan and {Kirby}, Evan and {Lupton}, Robert and {Peter}, Annika H.~G. and {Romanowsky}, Aaron J. and {Strader}, Jay},
        title = "{First Results from the MADCASH Survey: A Faint Dwarf Galaxy Companion to the Low-mass Spiral Galaxy NGC 2403 at 3.2 Mpc}",
      journal = {\apjl},
         year = 2016,
        month = sep,
       volume = {828},
       number = {1},
          eid = {L5},
        pages = {L5},
          doi = {10.3847/2041-8205/828/1/L5},
archivePrefix = {arXiv},
       eprint = {1608.02591},
 primaryClass = {astro-ph.GA},
       adsurl = {https://ui.adsabs.harvard.edu/abs/2016ApJ...828L...5C}
}

@ARTICLE{Carlin2019,
       author = {{Carlin}, Jeffrey L. and {Garling}, Christopher T. and {Peter}, Annika H.~G. and {Crnojevi{\'c}}, Denija and {Forbes}, Duncan A. and {Hargis}, Jonathan R. and {Mutlu-Pakdil}, Bur{\c{c}}in and {Pucha}, Ragadeepika and {Romanowsky}, Aaron J. and {Sand}, David J. and {Spekkens}, Kristine and {Strader}, Jay and {Willman}, Beth},
        title = "{Tidal Destruction in a Low-mass Galaxy Environment: The Discovery of Tidal Tails around DDO 44}",
      journal = {\apj},
         year = 2019,
        month = dec,
       volume = {886},
       number = {2},
          eid = {109},
        pages = {109},
          doi = {10.3847/1538-4357/ab4c32},
archivePrefix = {arXiv},
       eprint = {1906.08260},
 primaryClass = {astro-ph.GA},
       adsurl = {https://ui.adsabs.harvard.edu/abs/2019ApJ...886..109C}
}

@ARTICLE{Springel2001,
       author = {{Springel}, Volker and {White}, Simon D.~M. and {Tormen}, Giuseppe and {Kauffmann}, Guinevere},
        title = "{Populating a cluster of galaxies - I. Results at z=0}",
      journal = {\mnras},
         year = 2001,
        month = dec,
       volume = {328},
       number = {3},
        pages = {726-750},
          doi = {10.1046/j.1365-8711.2001.04912.x},
archivePrefix = {arXiv},
       eprint = {astro-ph/0012055},
 primaryClass = {astro-ph},
       adsurl = {https://ui.adsabs.harvard.edu/abs/2001MNRAS.328..726S}
}

@ARTICLE{Nelson2019,
       author = {{Nelson}, Dylan and {Pillepich}, Annalisa and {Springel}, Volker and {Pakmor}, R{\"u}diger and {Weinberger}, Rainer and {Genel}, Shy and {Torrey}, Paul and {Vogelsberger}, Mark and {Marinacci}, Federico and {Hernquist}, Lars},
        title = "{First results from the TNG50 simulation: galactic outflows driven by supernovae and black hole feedback}",
      journal = {\mnras},
         year = 2019,
        month = dec,
       volume = {490},
       number = {3},
        pages = {3234-3261},
          doi = {10.1093/mnras/stz2306},
archivePrefix = {arXiv},
       eprint = {1902.05554},
 primaryClass = {astro-ph.GA},
       adsurl = {https://ui.adsabs.harvard.edu/abs/2019MNRAS.490.3234N}
}

@ARTICLE{Nadler2020,
       author = {{Nadler}, E.~O. and {Wechsler}, R.~H. and {Bechtol}, K. and {Mao}, Y.-Y. and {Green}, G. and {Drlica-Wagner}, A. and {McNanna}, M. and {Mau}, S. and {Pace}, A.~B. and {Simon}, J.~D. and {Kravtsov}, A. and {Dodelson}, S. and {Li}, T.~S. and {Riley}, A.~H. and {Wang}, M.~Y. and {Abbott}, T.~M.~C. and {Aguena}, M. and {Allam}, S. and {Annis}, J. and {Avila}, S. and {Bernstein}, G.~M. and {Bertin}, E. and {Brooks}, D. and {Burke}, D.~L. and {Rosell}, A. Carnero and {Kind}, M. Carrasco and {Carretero}, J. and {Costanzi}, M. and {da Costa}, L.~N. and {De Vicente}, J. and {Desai}, S. and {Evrard}, A.~E. and {Flaugher}, B. and {Fosalba}, P. and {Frieman}, J. and {Garc{\'\i}a-Bellido}, J. and {Gaztanaga}, E. and {Gerdes}, D.~W. and {Gruen}, D. and {Gschwend}, J. and {Gutierrez}, G. and {Hartley}, W.~G. and {Hinton}, S.~R. and {Honscheid}, K. and {Krause}, E. and {Kuehn}, K. and {Kuropatkin}, N. and {Lahav}, O. and {Maia}, M.~A.~G. and {Marshall}, J.~L. and {Menanteau}, F. and {Miquel}, R. and {Palmese}, A. and {Paz-Chinch{\'o}n}, F. and {Plazas}, A.~A. and {Romer}, A.~K. and {Sanchez}, E. and {Santiago}, B. and {Scarpine}, V. and {Serrano}, S. and {Smith}, M. and {Soares-Santos}, M. and {Suchyta}, E. and {Tarle}, G. and {Thomas}, D. and {Varga}, T.~N. and {Walker}, A.~R. and {DES Collaboration}},
        title = "{Milky Way Satellite Census. II. Galaxy-Halo Connection Constraints Including the Impact of the Large Magellanic Cloud}",
      journal = {\apj},
         year = 2020,
        month = apr,
       volume = {893},
       number = {1},
          eid = {48},
        pages = {48},
          doi = {10.3847/1538-4357/ab846a},
archivePrefix = {arXiv},
       eprint = {1912.03303},
 primaryClass = {astro-ph.GA},
       adsurl = {https://ui.adsabs.harvard.edu/abs/2020ApJ...893...48N}
}

@ARTICLE{Taibi2025,
       author = {{Taibi}, S. and {Pawlowski}, M.~S. and {M{\"u}ller}, O. and {B{\'\i}lek}, M. and {J{\'u}lio}, M.~P. and {Kanehisa}, K.~J. and {Jovanovi{\'c}}, M. and {Lalovi{\'c}}, A. and {Samurovi{\'c}}, S.},
        title = "{An LBT view of the co-rotating group of galaxies around NGC 2750: Deep imaging and new satellite candidates}",
      journal = {\aap},
         year = 2025,
        month = jul,
       volume = {699},
          eid = {A285},
        pages = {A285},
          doi = {10.1051/0004-6361/202554365},
archivePrefix = {arXiv},
       eprint = {2506.19001},
 primaryClass = {astro-ph.GA},
       adsurl = {https://ui.adsabs.harvard.edu/abs/2025A&A...699A.285T}
}

@ARTICLE{Kanehisa2024,
       author = {{Kanehisa}, Kosuke Jamie and {Pawlowski}, Marcel S. and {Heesters}, Nick and {M{\"u}ller}, Oliver},
        title = "{A too-many dwarf satellite galaxies problem in the MATLAS low-to-moderate density fields}",
      journal = {\aap},
         year = 2024,
        month = jun,
       volume = {686},
          eid = {A280},
        pages = {A280},
          doi = {10.1051/0004-6361/202348242},
archivePrefix = {arXiv},
       eprint = {2405.05303},
 primaryClass = {astro-ph.GA},
       adsurl = {https://ui.adsabs.harvard.edu/abs/2024A&A...686A.280K}
}

@ARTICLE{Homma2024,
       author = {{Homma}, Daisuke and {Chiba}, Masashi and {Komiyama}, Yutaka and {Tanaka}, Masayuki and {Okamoto}, Sakurako and {Tanaka}, Mikito and {Ishigaki}, Miho N. and {Hayashi}, Kohei and {Arimoto}, Nobuo and {Lupton}, Robert H. and {Strauss}, Michael A. and {Miyazaki}, Satoshi and {Wang}, Shiang-Yu and {Murayama}, Hitoshi},
        title = "{Final results of the search for new Milky Way satellites in the Hyper Suprime-Cam Subaru Strategic Program survey: Discovery of two more candidates}",
      journal = {\pasj},
         year = 2024,
        month = aug,
       volume = {76},
       number = {4},
        pages = {733-752},
          doi = {10.1093/pasj/psae044},
archivePrefix = {arXiv},
       eprint = {2311.05439},
 primaryClass = {astro-ph.GA},
       adsurl = {https://ui.adsabs.harvard.edu/abs/2024PASJ...76..733H}
}

@ARTICLE{Muller2024,
       author = {{M{\"u}ller}, Oliver and {Pawlowski}, Marcel S. and {Revaz}, Yves and {Venhola}, Aku and {Rejkuba}, Marina and {Hilker}, Michael and {Lutz}, Katharina},
        title = "{A too-many-dwarf-galaxy-satellites problem in the M 83 group}",
      journal = {\aap},
         year = 2024,
        month = apr,
       volume = {684},
          eid = {L6},
        pages = {L6},
          doi = {10.1051/0004-6361/202348969},
archivePrefix = {arXiv},
       eprint = {2403.08717},
 primaryClass = {astro-ph.GA},
       adsurl = {https://ui.adsabs.harvard.edu/abs/2024A&A...684L...6M}
}

@ARTICLE{Li2026a,
       author = {{Li}, Jiaxuan and {Greene}, Jenny E. and {Danieli}, Shany and {Carlsten}, Scott G. and {Geha}, Marla},
        title = "{A Possible ``Too-many-satellites'' Problem in the Isolated Dwarf Galaxy DDO 161}",
      journal = {\apjl},
         year = 2026,
        month = feb,
       volume = {998},
       number = {1},
          eid = {L24},
        pages = {L24},
          doi = {10.3847/2041-8213/ae3ddd},
archivePrefix = {arXiv},
       eprint = {2511.01733},
 primaryClass = {astro-ph.GA},
       adsurl = {https://ui.adsabs.harvard.edu/abs/2026ApJ...998L..24L}
}

@ARTICLE{Diemer2021,
       author = {{Diemer}, Benedikt},
        title = "{Flybys, Orbits, Splashback: Subhalos and the Importance of the Halo Boundary}",
      journal = {\apj},
         year = 2021,
        month = mar,
       volume = {909},
       number = {2},
          eid = {112},
        pages = {112},
          doi = {10.3847/1538-4357/abd947},
archivePrefix = {arXiv},
       eprint = {2007.10992},
 primaryClass = {astro-ph.CO},
       adsurl = {https://ui.adsabs.harvard.edu/abs/2021ApJ...909..112D}
}

@article{Buck2019,
    author = {Buck, Tobias and Macciò, Andrea V and Dutton, Aaron A and Obreja, Aura and Frings, Jonas},
    title = {NIHAO XV: the environmental impact of the host galaxy on galactic satellite and field dwarf galaxies},
    journal = {Monthly Notices of the Royal Astronomical Society},
    volume = {483},
    number = {1},
    pages = {1314-1341},
    year = {2019},
    month = {02},
    issn = {0035-8711},
    doi = {10.1093/mnras/sty2913},
    url = {https://doi.org/10.1093/mnras/sty2913},
    eprint = {https://academic.oup.com/mnras/article-pdf/483/1/1314/27090431/sty2913.pdf},
}

@ARTICLE{Alpaslan2020,
       author = {{Alpaslan}, Mehmet and {Tinker}, Jeremy L.},
        title = "{Connecting SDSS central galaxies to their host haloes using total satellite luminosity}",
      journal = {\mnras},
         year = 2020,
        month = aug,
       volume = {496},
       number = {4},
        pages = {5463-5481},
          doi = {10.1093/mnras/staa1844},
archivePrefix = {arXiv},
       eprint = {1911.04509},
 primaryClass = {astro-ph.GA},
       adsurl = {https://ui.adsabs.harvard.edu/abs/2020MNRAS.496.5463A}
}

@ARTICLE{Wang2014,
       author = {{Wang}, Wenting and {Sales}, Laura V. and {Henriques}, Bruno M.~B. and {White}, Simon D.~M.},
        title = "{Satellite abundances around bright isolated galaxies - II. Radial distribution and environmental effects}",
      journal = {\mnras},
         year = 2014,
        month = aug,
       volume = {442},
       number = {2},
        pages = {1363-1378},
          doi = {10.1093/mnras/stu988},
archivePrefix = {arXiv},
       eprint = {1403.2409},
 primaryClass = {astro-ph.CO},
       adsurl = {https://ui.adsabs.harvard.edu/abs/2014MNRAS.442.1363W}
}

@ARTICLE{Chen2006,
       author = {{Chen}, Jacqueline and {Kravtsov}, Andrey V. and {Prada}, Francisco and {Sheldon}, Erin S. and {Klypin}, Anatoly A. and {Blanton}, Michael R. and {Brinkmann}, Jonathan and {Thakar}, Aniruddha R.},
        title = "{Constraining the Projected Radial Distribution of Galactic Satellites with the Sloan Digital Sky Survey}",
      journal = {\apj},
         year = 2006,
        month = aug,
       volume = {647},
       number = {1},
        pages = {86-101},
          doi = {10.1086/504462},
archivePrefix = {arXiv},
       eprint = {astro-ph/0512376},
 primaryClass = {astro-ph},
       adsurl = {https://ui.adsabs.harvard.edu/abs/2006ApJ...647...86C}
}

@ARTICLE{White1978,
       author = {{White}, S.~D.~M. and {Rees}, M.~J.},
        title = "{Core condensation in heavy halos: a two-stage theory for galaxy formation and clustering.}",
      journal = {\mnras},
         year = 1978,
        month = may,
       volume = {183},
        pages = {341-358},
          doi = {10.1093/mnras/183.3.341},
       adsurl = {https://ui.adsabs.harvard.edu/abs/1978MNRAS.183..341W}
}

@ARTICLE{Bryan1998,
       author = {{Bryan}, Greg L. and {Norman}, Michael L.},
        title = "{Statistical Properties of X-Ray Clusters: Analytic and Numerical Comparisons}",
      journal = {\apj},
         year = 1998,
        month = mar,
       volume = {495},
       number = {1},
        pages = {80-99},
          doi = {10.1086/305262},
archivePrefix = {arXiv},
       eprint = {astro-ph/9710107},
 primaryClass = {astro-ph},
       adsurl = {https://ui.adsabs.harvard.edu/abs/1998ApJ...495...80B}
}

@ARTICLE{Spergel2000,
       author = {{Spergel}, David N. and {Steinhardt}, Paul J.},
        title = "{Observational Evidence for Self-Interacting Cold Dark Matter}",
      journal = {\prl},
         year = 2000,
        month = apr,
       volume = {84},
       number = {17},
        pages = {3760-3763},
          doi = {10.1103/PhysRevLett.84.3760},
archivePrefix = {arXiv},
       eprint = {astro-ph/9909386},
 primaryClass = {astro-ph},
       adsurl = {https://ui.adsabs.harvard.edu/abs/2000PhRvL..84.3760S}
}

@ARTICLE{Pan2026,
       author = {{Pan}, Yue and {Danieli}, Shany and {Greene}, Jenny E. and {Li}, Jiaxuan and {Leauthaud}, Alexie and {Kado-Fong}, Erin and {Luo}, Yifei and {Mintz}, Abby and {Brooks}, Alyson and {Huang}, Song and {Peter}, Annika H.~G. and {Bhattacharyya}, Joy and {Kelvin}, Lee S.},
        title = "{The Merian Survey: A Statistical Census of Bright Satellites of Milky Way Analogs}",
      journal = {\apj},
         year = 2026,
        month = mar,
       volume = {1000},
       number = {1},
          eid = {119},
        pages = {119},
          doi = {10.3847/1538-4357/ae42cb},
archivePrefix = {arXiv},
       eprint = {2512.12846},
 primaryClass = {astro-ph.GA},
       adsurl = {https://ui.adsabs.harvard.edu/abs/2026ApJ..1000..119P}
}

@ARTICLE{Li2025,
       author = {{Li}, Jiaxuan and {Greene}, Jenny E. and {Danieli}, Shany and {Carlsten}, Scott G. and {Geha}, Marla and {Jiang}, Fangzhou and {Tanaka}, Masayuki},
        title = "{ELVES-Dwarf I: Satellites Systems of Eight Isolated Dwarf Galaxies in the Local Volume}",
      journal = {arXiv e-prints},
         year = 2025,
        month = apr,
          eid = {arXiv:2504.08030},
        pages = {arXiv:2504.08030},
          doi = {10.48550/arXiv.2504.08030},
archivePrefix = {arXiv},
       eprint = {2504.08030},
 primaryClass = {astro-ph.GA},
       adsurl = {https://ui.adsabs.harvard.edu/abs/2025arXiv250408030L}
}

@ARTICLE{Falugher2015,
       author = {{Flaugher}, B. and {Diehl}, H.~T. and {Honscheid}, K. and {Abbott}, T.~M.~C. and {Alvarez}, O. and {Angstadt}, R. and {Annis}, J.~T. and {Antonik}, M. and {Ballester}, O. and {Beaufore}, L. and {Bernstein}, G.~M. and {Bernstein}, R.~A. and {Bigelow}, B. and {Bonati}, M. and {Boprie}, D. and {Brooks}, D. and {Buckley-Geer}, E.~J. and {Campa}, J. and {Cardiel-Sas}, L. and {Castander}, F.~J. and {Castilla}, J. and {Cease}, H. and {Cela-Ruiz}, J.~M. and {Chappa}, S. and {Chi}, E. and {Cooper}, C. and {da Costa}, L.~N. and {Dede}, E. and {Derylo}, G. and {DePoy}, D.~L. and {de Vicente}, J. and {Doel}, P. and {Drlica-Wagner}, A. and {Eiting}, J. and {Elliott}, A.~E. and {Emes}, J. and {Estrada}, J. and {Fausti Neto}, A. and {Finley}, D.~A. and {Flores}, R. and {Frieman}, J. and {Gerdes}, D. and {Gladders}, M.~D. and {Gregory}, B. and {Gutierrez}, G.~R. and {Hao}, J. and {Holland}, S.~E. and {Holm}, S. and {Huffman}, D. and {Jackson}, C. and {James}, D.~J. and {Jonas}, M. and {Karcher}, A. and {Karliner}, I. and {Kent}, S. and {Kessler}, R. and {Kozlovsky}, M. and {Kron}, R.~G. and {Kubik}, D. and {Kuehn}, K. and {Kuhlmann}, S. and {Kuk}, K. and {Lahav}, O. and {Lathrop}, A. and {Lee}, J. and {Levi}, M.~E. and {Lewis}, P. and {Li}, T.~S. and {Mandrichenko}, I. and {Marshall}, J.~L. and {Martinez}, G. and {Merritt}, K.~W. and {Miquel}, R. and {Mu{\~n}oz}, F. and {Neilsen}, E.~H. and {Nichol}, R.~C. and {Nord}, B. and {Ogando}, R. and {Olsen}, J. and {Palaio}, N. and {Patton}, K. and {Peoples}, J. and {Plazas}, A.~A. and {Rauch}, J. and {Reil}, K. and {Rheault}, J. -P. and {Roe}, N.~A. and {Rogers}, H. and {Roodman}, A. and {Sanchez}, E. and {Scarpine}, V. and {Schindler}, R.~H. and {Schmidt}, R. and {Schmitt}, R. and {Schubnell}, M. and {Schultz}, K. and {Schurter}, P. and {Scott}, L. and {Serrano}, S. and {Shaw}, T.~M. and {Smith}, R.~C. and {Soares-Santos}, M. and {Stefanik}, A. and {Stuermer}, W. and {Suchyta}, E. and {Sypniewski}, A. and {Tarle}, G. and {Thaler}, J. and {Tighe}, R. and {Tran}, C. and {Tucker}, D. and {Walker}, A.~R. and {Wang}, G. and {Watson}, M. and {Weaverdyck}, C. and {Wester}, W. and {Woods}, R. and {Yanny}, B. and {DES Collaboration}},
        title = "{The Dark Energy Camera}",
      journal = {\aj},
         year = 2015,
        month = nov,
       volume = {150},
       number = {5},
          eid = {150},
        pages = {150},
          doi = {10.1088/0004-6256/150/5/150},
archivePrefix = {arXiv},
       eprint = {1504.02900},
 primaryClass = {astro-ph.IM},
       adsurl = {https://ui.adsabs.harvard.edu/abs/2015AJ....150..150F}
}

@ARTICLE{Dooley2017,
       author = {{Dooley}, Gregory A. and {Peter}, Annika H.~G. and {Yang}, Tianyi and {Willman}, Beth and {Griffen}, Brendan F. and {Frebel}, Anna},
        title = "{An observer's guide to the (Local Group) dwarf galaxies: predictions for their own dwarf satellite populations}",
      journal = {\mnras},
         year = 2017,
        month = nov,
       volume = {471},
       number = {4},
        pages = {4894-4909},
          doi = {10.1093/mnras/stx1900},
archivePrefix = {arXiv},
       eprint = {1610.00708},
 primaryClass = {astro-ph.GA},
       adsurl = {https://ui.adsabs.harvard.edu/abs/2017MNRAS.471.4894D}
}

@ARTICLE{Santos2022,
       author = {{Santos-Santos}, Isabel M.~E. and {Sales}, Laura V. and {Fattahi}, Azadeh and {Navarro}, Julio F.},
        title = "{Satellite mass functions and the faint end of the galaxy mass-halo mass relation in LCDM}",
      journal = {\mnras},
         year = 2022,
        month = sep,
       volume = {515},
       number = {3},
        pages = {3685-3697},
          doi = {10.1093/mnras/stac2057},
archivePrefix = {arXiv},
       eprint = {2111.01158},
 primaryClass = {astro-ph.GA},
       adsurl = {https://ui.adsabs.harvard.edu/abs/2022MNRAS.515.3685S}
}

@ARTICLE{Savino2025,
       author = {{Savino}, Alessandro and {Weisz}, Daniel R. and {Dolphin}, Andrew E. and {Durbin}, Meredith J. and {Kallivayalil}, Nitya and {Wetzel}, Andrew and {Anderson}, Jay and {Besla}, Gurtina and {Boylan-Kolchin}, Michael and {Brown}, Thomas M. and {Bullock}, James S. and {Cole}, Andrew A. and {Collins}, Michelle L.~M. and {Cooper}, M.~C. and {Deason}, Alis J. and {Dotter}, Aaron L. and {Fardal}, Mark and {Ferguson}, Annette M.~N. and {Fritz}, Tobias K. and {Geha}, Marla C. and {Gilbert}, Karoline M. and {Guhathakurta}, Puragra and {Ibata}, Rodrigo and {Irwin}, Michael J. and {Jeon}, Myoungwon and {Kirby}, Evan N. and {Lewis}, Geraint F. and {Mackey}, Dougal and {Majewski}, Steven R. and {Martin}, Nicolas and {McConnachie}, Alan and {Patel}, Ekta and {Rich}, R. Michael and {Skillman}, Evan D. and {Simon}, Joshua D. and {Sohn}, Sangmo Tony and {Tollerud}, Erik J. and {van der Marel}, Roeland P.},
        title = "{The Hubble Space Telescope Survey of M31 Satellite Galaxies. IV. Survey Overview and Lifetime Star Formation Histories}",
      journal = {\apj},
         year = 2025,
        month = feb,
       volume = {979},
       number = {2},
          eid = {205},
        pages = {205},
          doi = {10.3847/1538-4357/ada24f},
archivePrefix = {arXiv},
       eprint = {2501.13152},
 primaryClass = {astro-ph.GA},
       adsurl = {https://ui.adsabs.harvard.edu/abs/2025ApJ...979..205S}
}

@ARTICLE{Hopkins2011,
       author = {{Hopkins}, Philip F. and {Quataert}, Eliot and {Murray}, Norman},
        title = "{Self-regulated star formation in galaxies via momentum input from massive stars}",
      journal = {\mnras},
         year = 2011,
        month = oct,
       volume = {417},
       number = {2},
        pages = {950-973},
          doi = {10.1111/j.1365-2966.2011.19306.x},
archivePrefix = {arXiv},
       eprint = {1101.4940},
 primaryClass = {astro-ph.CO},
       adsurl = {https://ui.adsabs.harvard.edu/abs/2011MNRAS.417..950H}
}

@ARTICLE{Gemmel2024,
       author = {{Gemmell}, Caleb and {Roy}, Sandip and {Shen}, Xuejian and {Curtin}, David and {Lisanti}, Mariangela and {Murray}, Norman and {Hopkins}, Philip F.},
        title = "{Dissipative Dark Substructure: The Consequences of Atomic Dark Matter on Milky Way Analog Subhalos}",
      journal = {\apj},
         year = 2024,
        month = may,
       volume = {967},
       number = {1},
          eid = {21},
        pages = {21},
          doi = {10.3847/1538-4357/ad3823},
archivePrefix = {arXiv},
       eprint = {2311.02148},
 primaryClass = {astro-ph.GA},
       adsurl = {https://ui.adsabs.harvard.edu/abs/2024ApJ...967...21G}
}

@ARTICLE{Bullock2017,
       author = {{Bullock}, James S. and {Boylan-Kolchin}, Michael},
        title = "{Small-Scale Challenges to the {\ensuremath{\Lambda}}CDM Paradigm}",
      journal = {\araa},
         year = 2017,
        month = aug,
       volume = {55},
       number = {1},
        pages = {343-387},
          doi = {10.1146/annurev-astro-091916-055313},
archivePrefix = {arXiv},
       eprint = {1707.04256},
 primaryClass = {astro-ph.CO},
       adsurl = {https://ui.adsabs.harvard.edu/abs/2017ARA&A..55..343B}
}

@ARTICLE{Tinker2017,
       author = {{Tinker}, Jeremy L. and {Brownstein}, Joel R. and {Guo}, Hong and {Leauthaud}, Alexie and {Maraston}, Claudia and {Masters}, Karen and {Montero-Dorta}, Antonio D. and {Thomas}, Daniel and {Tojeiro}, Rita and {Weiner}, Benjamin and {Zehavi}, Idit and {Olmstead}, Matthew D.},
        title = "{The Correlation between Halo Mass and Stellar Mass for the Most Massive Galaxies in the Universe}",
      journal = {\apj},
         year = 2017,
        month = apr,
       volume = {839},
       number = {2},
          eid = {121},
        pages = {121},
          doi = {10.3847/1538-4357/aa6845},
archivePrefix = {arXiv},
       eprint = {1607.04678},
 primaryClass = {astro-ph.GA},
       adsurl = {https://ui.adsabs.harvard.edu/abs/2017ApJ...839..121T}
}

@ARTICLE{Rodriguez2017,
       author = {{Rodr{\'\i}guez-Puebla}, Aldo and {Primack}, Joel R. and {Avila-Reese}, Vladimir and {Faber}, S.~M.},
        title = "{Constraining the galaxy-halo connection over the last 13.3 Gyr: star formation histories, galaxy mergers and structural properties}",
      journal = {\mnras},
         year = 2017,
        month = sep,
       volume = {470},
       number = {1},
        pages = {651-687},
          doi = {10.1093/mnras/stx1172},
archivePrefix = {arXiv},
       eprint = {1703.04542},
 primaryClass = {astro-ph.GA},
       adsurl = {https://ui.adsabs.harvard.edu/abs/2017MNRAS.470..651R}
}

@ARTICLE{2018AJ....156..123A,
       author = {{Astropy Collaboration} and {Price-Whelan}, A.~M. and {Sip{\H{o}}cz}, B.~M. and {G{\"u}nther}, H.~M. and {Lim}, P.~L. and {Crawford}, S.~M. and {Conseil}, S. and {Shupe}, D.~L. and {Craig}, M.~W. and {Dencheva}, N. and {Ginsburg}, A. and {VanderPlas}, J.~T. and {Bradley}, L.~D. and {P{\'e}rez-Su{\'a}rez}, D. and {de Val-Borro}, M. and {Aldcroft}, T.~L. and {Cruz}, K.~L. and {Robitaille}, T.~P. and {Tollerud}, E.~J. and {Ardelean}, C. and {Babej}, T. and {Bach}, Y.~P. and {Bachetti}, M. and {Bakanov}, A.~V. and {Bamford}, S.~P. and {Barentsen}, G. and {Barmby}, P. and {Baumbach}, A. and {Berry}, K.~L. and {Biscani}, F. and {Boquien}, M. and {Bostroem}, K.~A. and {Bouma}, L.~G. and {Brammer}, G.~B. and {Bray}, E.~M. and {Breytenbach}, H. and {Buddelmeijer}, H. and {Burke}, D.~J. and {Calderone}, G. and {Cano Rodr{\'\i}guez}, J.~L. and {Cara}, M. and {Cardoso}, J.~V.~M. and {Cheedella}, S. and {Copin}, Y. and {Corrales}, L. and {Crichton}, D. and {D'Avella}, D. and {Deil}, C. and {Depagne}, {\'E}. and {Dietrich}, J.~P. and {Donath}, A. and {Droettboom}, M. and {Earl}, N. and {Erben}, T. and {Fabbro}, S. and {Ferreira}, L.~A. and {Finethy}, T. and {Fox}, R.~T. and {Garrison}, L.~H. and {Gibbons}, S.~L.~J. and {Goldstein}, D.~A. and {Gommers}, R. and {Greco}, J.~P. and {Greenfield}, P. and {Groener}, A.~M. and {Grollier}, F. and {Hagen}, A. and {Hirst}, P. and {Homeier}, D. and {Horton}, A.~J. and {Hosseinzadeh}, G. and {Hu}, L. and {Hunkeler}, J.~S. and {Ivezi{\'c}}, {\v{Z}}. and {Jain}, A. and {Jenness}, T. and {Kanarek}, G. and {Kendrew}, S. and {Kern}, N.~S. and {Kerzendorf}, W.~E. and {Khvalko}, A. and {King}, J. and {Kirkby}, D. and {Kulkarni}, A.~M. and {Kumar}, A. and {Lee}, A. and {Lenz}, D. and {Littlefair}, S.~P. and {Ma}, Z. and {Macleod}, D.~M. and {Mastropietro}, M. and {McCully}, C. and {Montagnac}, S. and {Morris}, B.~M. and {Mueller}, M. and {Mumford}, S.~J. and {Muna}, D. and {Murphy}, N.~A. and {Nelson}, S. and {Nguyen}, G.~H. and {Ninan}, J.~P. and {N{\"o}the}, M. and {Ogaz}, S. and {Oh}, S. and {Parejko}, J.~K. and {Parley}, N. and {Pascual}, S. and {Patil}, R. and {Patil}, A.~A. and {Plunkett}, A.~L. and {Prochaska}, J.~X. and {Rastogi}, T. and {Reddy Janga}, V. and {Sabater}, J. and {Sakurikar}, P. and {Seifert}, M. and {Sherbert}, L.~E. and {Sherwood-Taylor}, H. and {Shih}, A.~Y. and {Sick}, J. and {Silbiger}, M.~T. and {Singanamalla}, S. and {Singer}, L.~P. and {Sladen}, P.~H. and {Sooley}, K.~A. and {Sornarajah}, S. and {Streicher}, O. and {Teuben}, P. and {Thomas}, S.~W. and {Tremblay}, G.~R. and {Turner}, J.~E.~H. and {Terr{\'o}n}, V. and {van Kerkwijk}, M.~H. and {de la Vega}, A. and {Watkins}, L.~L. and {Weaver}, B.~A. and {Whitmore}, J.~B. and {Woillez}, J. and {Zabalza}, V. and {Astropy Contributors}},
        title = "{The Astropy Project: Building an Open-science Project and Status of the v2.0 Core Package}",
      journal = {\aj},
         year = 2018,
        month = sep,
       volume = {156},
       number = {3},
          eid = {123},
        pages = {123},
          doi = {10.3847/1538-3881/aabc4f},
archivePrefix = {arXiv},
       eprint = {1801.02634},
 primaryClass = {astro-ph.IM},
       adsurl = {https://ui.adsabs.harvard.edu/abs/2018AJ....156..123A}
}

@ARTICLE{2013A&A...558A..33A,
       author = {{Astropy Collaboration} and {Robitaille}, Thomas P. and
         {Tollerud}, Erik J. and {Greenfield}, Perry and {Droettboom}, Michael and
         {Bray}, Erik and {Aldcroft}, Tom and {Davis}, Matt and
         {Ginsburg}, Adam and {Price-Whelan}, Adrian M. and
         {Kerzendorf}, Wolfgang E. and {Conley}, Alexander and {Crighton}, Neil and
         {Barbary}, Kyle and {Muna}, Demitri and {Ferguson}, Henry and
         {Grollier}, Fr{\'e}d{\'e}ric and {Parikh}, Madhura M. and
         {Nair}, Prasanth H. and {Unther}, Hans M. and {Deil}, Christoph and
         {Woillez}, Julien and {Conseil}, Simon and {Kramer}, Roban and
         {Turner}, James E.~H. and {Singer}, Leo and {Fox}, Ryan and
         {Weaver}, Benjamin A. and {Zabalza}, Victor and {Edwards}, Zachary I. and
         {Azalee Bostroem}, K. and {Burke}, D.~J. and {Casey}, Andrew R. and
         {Crawford}, Steven M. and {Dencheva}, Nadia and {Ely}, Justin and
         {Jenness}, Tim and {Labrie}, Kathleen and {Lim}, Pey Lian and
         {Pierfederici}, Francesco and {Pontzen}, Andrew and {Ptak}, Andy and
         {Refsdal}, Brian and {Servillat}, Mathieu and {Streicher}, Ole},
        title = "{Astropy: A community Python package for astronomy}",
      journal = {\aap},
         year = "2013",
        month = "Oct",
       volume = {558},
          eid = {A33},
        pages = {A33},
          doi = {10.1051/0004-6361/201322068},
archivePrefix = {arXiv},
       eprint = {1307.6212},
 primaryClass = {astro-ph.IM},
       adsurl = {https://ui.adsabs.harvard.edu/abs/2013A&A...558A..33A}
}

@ARTICLE{Behroozi.etal.2010,
       author = {{Behroozi}, Peter S. and {Conroy}, Charlie and {Wechsler}, Risa H.},
        title = "{A Comprehensive Analysis of Uncertainties Affecting the Stellar Mass-Halo Mass Relation for 0 < z < 4}",
      journal = {\apj},
         year = 2010,
        month = jul,
       volume = {717},
       number = {1},
        pages = {379-403},
          doi = {10.1088/0004-637X/717/1/379},
archivePrefix = {arXiv},
       eprint = {1001.0015},
 primaryClass = {astro-ph.CO},
       adsurl = {https://ui.adsabs.harvard.edu/abs/2010ApJ...717..379B}
}

@ARTICLE{Carlsten.etal.2021,
       author = {{Carlsten}, Scott G. and {Greene}, Jenny E. and {Greco}, Johnny P. and {Beaton}, Rachael L. and {Kado-Fong}, Erin},
        title = "{Structures of Dwarf Satellites of Milky Way-like Galaxies: Morphology, Scaling Relations, and Intrinsic Shapes}",
      journal = {\apj},
         year = 2021,
        month = dec,
       volume = {922},
       number = {2},
          eid = {267},
        pages = {267},
          doi = {10.3847/1538-4357/ac2581},
archivePrefix = {arXiv},
       eprint = {2105.03435},
 primaryClass = {astro-ph.GA},
       adsurl = {https://ui.adsabs.harvard.edu/abs/2021ApJ...922..267C}
}
\bibliographystyle{aasjournalv7}

\appendix

\setcounter{equation}{0}
\setcounter{figure}{0}
\setcounter{table}{0}
\renewcommand{\theequation}{A\arabic{equation}}
\renewcommand{\thefigure}{A\arabic{figure}}
\renewcommand{\thetable}{A\arabic{table}}
\renewcommand{\theHequation}{A\arabic{equation}}
\renewcommand{\theHfigure}{A\arabic{figure}}
\renewcommand{\theHtable}{A\arabic{table}}

\section{Uncertainty in Virial Radius}\label{sec:rvir}

Given the uncertainty in the stellar mass of the hosts and the intrinsic scatter of the halo mass-stellar mass relation~\citep{Tinker2017}, we find an average uncertainty on the true virial radii of $R_{\rm vir,true}/R_{\text{vir}}=1.0^{+0.3}_{-0.2}$ (where $R_{\text{vir}}$ is our computed value). To account for this, we extend the search for satellite candidates to $1.3R_{\text{vir}}$ of each host, assuming the same redshift. After visual inspection, we identify an additional $296$ satellite candidates, expanding the catalog to $760$. The increase in virial area universally increases the estimated background by $1.69\times$, bringing the background-subtracted sample to $214\pm 11$ satellites. 
\begin{figure*}[t!]
\centering
\includegraphics[width=0.46\columnwidth]{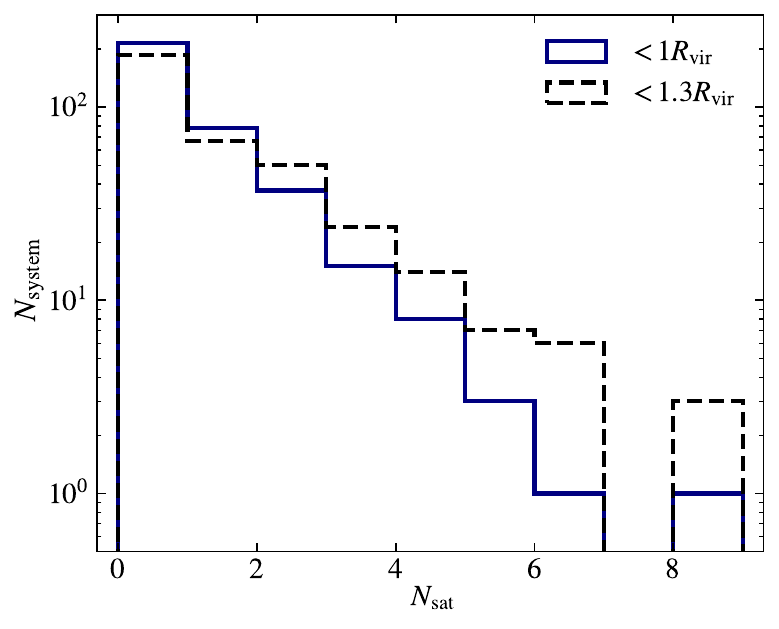}
\caption{Distribution of number of satellites across all hosts. The solid navy line represents the background-subtracted sample of satellites within $1R_{\rm vir}$ of their hosts. The dashed black line represents the background-subtracted sample with the search extended to $1.3R_{\rm vir}$.
\label{fig:Nsat_Nsys_13R}}
\end{figure*}

Figure~\ref{fig:Nsat_Nsys_13R} shows the satellite abundance distribution for the choices of $1R_{\rm vir}$ and $1.3R_{\rm vir}$, The latter increases the number of hosts near the high-$N_{\rm sat}$ tail, with $\mathord{\sim}15\%$ of hosts now having $N_{\rm sat}\geq 3$, compared to $\mathord{\sim} 8\%$ within $1R_{\rm vir}$. Furthermore, only $\mathord{\sim} 52\%$ of hosts have $N_{\rm sat}<1$. 

\begin{figure*}[t!]
\centering
\includegraphics[width=0.33\textwidth]{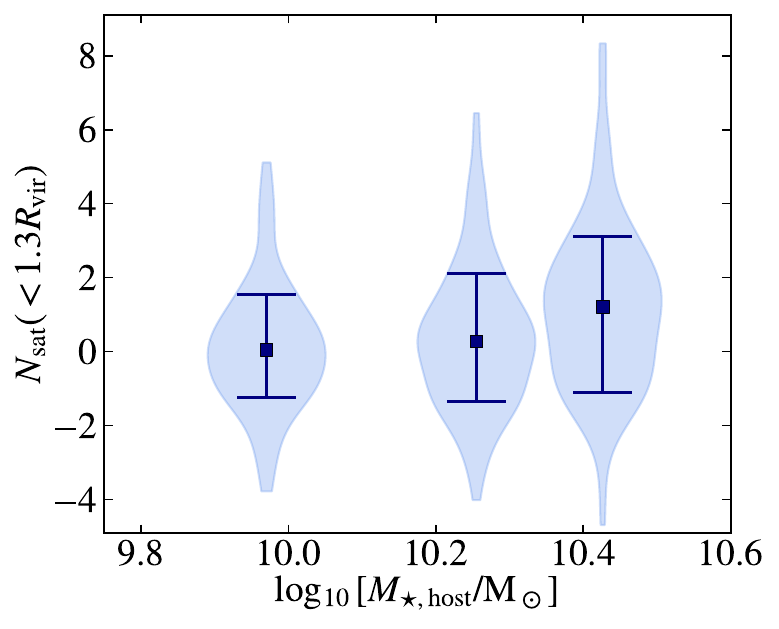}\includegraphics[width=0.33\textwidth]{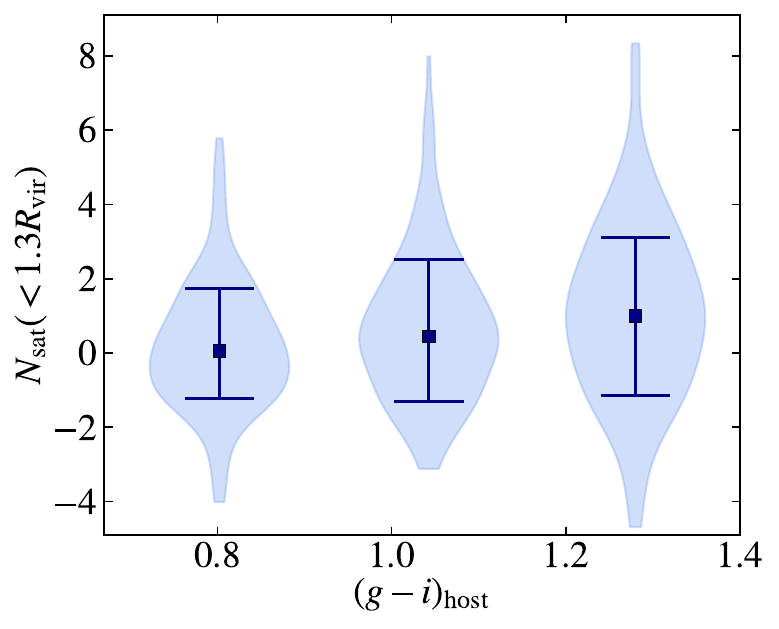}\includegraphics[width=0.33\textwidth]{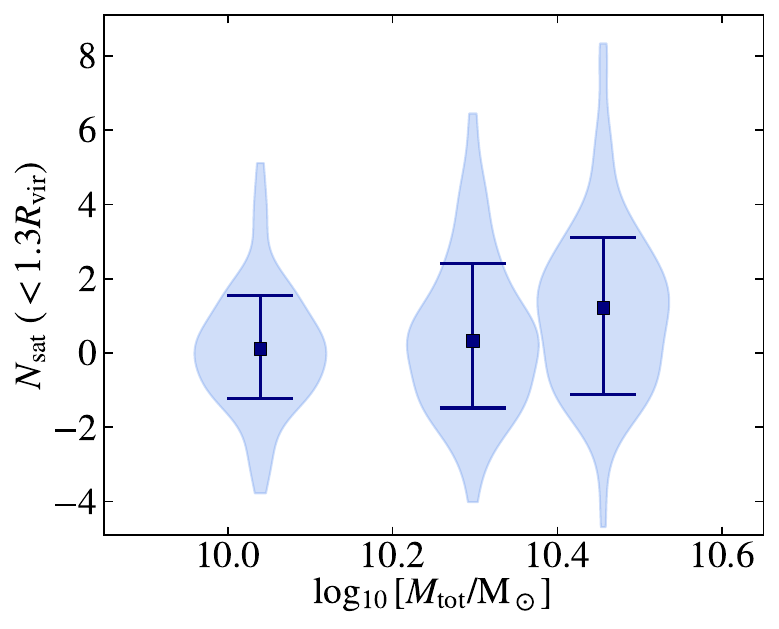}
\caption{\textbf{Left:} Satellite abundance $N_{\text{sat}}$ within $1.3R_{\rm vir}$ as a function of host stellar mass $M_{\star,\rm host}$. The blue bins correspond to: $M_\star < 10^{10.1} \ \rm M_\odot$, $10^{10.1} \ \text{M}_\odot \leq M_{\star, \rm host} < 10^{10.35} \ \rm M_\odot$, and $M_{\star, \rm host} \geq 10^{10.35} \ \rm M_\odot$. The median $N_{\text{sat}}$ in each bin is shown as a square with the $16^{\rm th}$ and $84^{\rm th}$ percentiles as horizontal bars. \textbf{Middle:} $N_{\text{sat}}$ as a function of host color $(g-i)_{\rm host}$. The blue bins correspond to: $(g-i)_{\rm host} < 0.9$, $0.9 \leq (g-i)_{\rm host} < 1.2$, and $(g-i)_{\rm host} \geq 1.2$. \textbf{Right:} $N_{\text{sat}}(<1.3R_{\rm vir})$ as a function of system total stellar mass $M_{\text{tot}}$. The blue bins correspond to: $M_{\rm tot} < 10^{10.15} \ \rm M_\odot$, $10^{10.15} \ \text{M}_\odot \leq M_{\rm tot} < 10^{10.35} \ \rm M_\odot$, and $M_{\rm tot} \geq 10^{10.35} \ \rm M_\odot$.
\label{fig:violins_13R}}
\end{figure*}

The left panel of Figure \ref{fig:violins_13R} shows the connection between $N_{\text{sat}}(<1.3R_{\rm vir})$ and $M_{\star,\text{host}}$. The hosts are divided into the same bins as in Figure~\ref{fig:violins}. The figure shows correlation between both median and $84^{\rm th}$ percentile of $N_{\text{sat}}$ and $M_{\star, \rm host}$, with the three bins having median and $68\%$ containment values of $0.0^{+1.6}_{-1.2}$, $0.3^{+1.8}_{-1.6}$, and $1.2^{+1.9}_{-2.3}$. The correlation coefficient for the medians is lower than for $N_{\text{sat}}(<R_{\rm vir})$, with $\rho=0.89$. This larger search radius increases the spread in the distribution, universally increasing the $84^{\rm th}$ percentiles due to the greater signal and decreasing the $16^{\rm th}$ percentiles due to the larger background. 

The middle panel of Figure \ref{fig:violins_13R} shows the connection between $N_{\text{sat}}(<1.3R_{\rm vir})$ and $(g-i)_{\rm host}$. The hosts are divided into the same bins as in Figure~\ref{fig:violins}. The figure shows correlation between both median and $84^{\rm th}$ percentile of $N_{\text{sat}}$ and $M_{\star, \rm host}$, with the three bins having median and $68\%$ containment values of $0.0^{+1.6}_{-1.2}$, $0.3^{+1.8}_{-1.6}$, and $1.2^{+1.9}_{-2.3}$. The correlation coefficient for the medians is lower than for $N_{\text{sat}}(<R_{\rm vir})$, with $\rho=0.93$.

Finally, the same trends hold for the right panel of Figure \ref{fig:violins_13R}, which shows the connection between $N_{\text{sat}}(<1.3R_{\rm vir})$ and $M_{\text{tot}}$. The three bins, equivalent to the ones in Figure~\ref{fig:violins}, have median and $68\%$ containment values of $-0.2^{+1.3}_{-1.1}$, $0.3^{+1.4}_{-1.9}$, and $1.2^{+1.8}_{-2.1}$. The correlation between $N_{\text{sat}}$ and both $M_{\star,\text{host}}$ and $M_{\text{tot}}$ within the sample are robust within the stated uncertainty in $R_{\rm vir}$. The correlation coefficient for the medians is lower than for $N_{\text{sat}}(<R_{\rm vir})$, with $\rho=0.88$.

These results illustrate how there are likely many more dwarf galaxies associated with the hosts beyond their predicted $R_{\rm vir}$. This increased signal may stem from inaccurate measurements of $R_{\rm vir}$ or from infalling or backsplash satellites beyond the accurate $R_{\rm vir}$~\citep{Buck2019,Diemer2021}. The correlation coefficient between $N_{\text{sat}}(<1.3R_{\rm vir})$ and the three host properties evaluated were lower than for $N_{\text{sat}}(<R_{\rm vir})$, indicating these outer regions may be less influenced by the host environment. 

\end{CJK*}
\end{document}